\documentclass[letterpaper]{article}
\usepackage[preprint]{aaai2027}
\usepackage[hyphens]{url}
\usepackage{graphicx}
\usepackage{natbib}
\usepackage{caption}
\usepackage{float}
\usepackage{booktabs}
\usepackage{amsmath,amssymb,amsthm}

\newcommand{\X}{\mathcal X}

\renewcommand{\vec}{\mathbf}
\newcommand{\Description}[1]{}

\title{Understanding Human Perception of Representation in Citizens' Assemblies: An Empirical Study}
\author{Yusuf Hakan Kalayci\textsuperscript{1},
Vasilis Varsamis\textsuperscript{1},
Nick Gill\textsuperscript{2},
Evi Micha\textsuperscript{1}}
\affiliations{\textsuperscript{1}\,University of Southern California\\
\textsuperscript{2}\,Sortition Foundation}

\begin{document}
\raggedbottom
\maketitle

\begin{abstract}

Citizens' assemblies are deliberative bodies intended to form a microcosm of the population. Organizers rely on quota-based stratification and must decide which attributes define resemblance to the public. Yet meeting every quota can still leave a dimension citizens value unrepresented. We study this attribute-selection problem in general-purpose and climate-focused assemblies through randomized conjoint experiments. Our main observation is that demographic attributes matter for perceived representation, but political alignment and context-specific attributes such as climate concern,  exert a stronger influence.
When both are shown in a climate-focused setting, each remains influential, with political alignment having the larger estimated marginal effect. We also examine what happens when a relevant attribute is omitted from stratification. Panels stratified on demographics, even with political alignment included, match the observed pool's climate-concern distribution no better than uniform random samples. These results suggest that representation on a relevant topic-specific attribute cannot always be recovered through correlated demographic or political quotas, and may therefore require explicit stratification. Finally, we ask whether representation preferences can be learned from the observed profiles. We study a hierarchy of predictive models, ranging from simple, interpretable matching rules to a learned metric representation and a respondent-conditioned utility model. The two learned models predict choices for respondents excluded from training with substantial accuracy, showing that representation judgments contain generalizable structure, although neither captures these judgments fully. Together, these findings offer empirical guidance for attribute selection in citizens' assemblies: designers should consider political and topic-specific dimensions alongside demographic attributes, avoid assuming that correlated proxy attributes will protect important omitted dimensions, and treat predictive models as informative diagnostics of how observed profiles shape representation choices.
\end{abstract}

\section{Introduction}
\label{sec:introduction}

Citizens' assemblies are increasingly used to bring members of the public together to deliberate on complex issues and inform decision-making across a wide range of policy and community contexts including constitutional reform and climate change policy
\citep{oecd2020deliberative,pilet2023public,curato2021deliberative}. Because a small panel is asked to
deliberate on behalf of a wider public, fairness and representativeness are
central design concerns. In the OECD's terms, random selection seeks to ensure
that ``nearly every person has an equal chance of being invited'' and that ``the
final group is a microcosm of society'' \citep{oecd2020deliberative}. In
practice, this resemblance is defined through self-reported attributes such as
gender, age, geography, and socioeconomic position.

The attributes above are  inputs to
quota-based stratification, the predominant approach to panel selection in
practice. Recent computational work improves fairness, transparency, and
robustness within the resulting feasible set
\citep{flanigan2021fair,flanigan2024transparent,assos2025alternates,
gambhir2025dropout}. However, these methods take the chosen attributes and
categories as given. 

These considerations raise three fundamental empirical questions for the design of citizens' assemblies. First, which attributes do people actually want to be represented on? Second, if a relevant attribute is omitted from the quota schema, are demographic and political attributes sufficient to capture the resulting heterogeneity, or must the omitted attribute be represented explicitly? Third, how do people combine multiple attributes when judging whether someone represents them? These questions matter in practice because no panel-selection design can represent every conceivable feature. As the set of potentially relevant attributes expands, the number of possible intersectional profiles grows combinatorially, while recruitment constraints, respondent burden, and limited panel sizes restrict the number of attributes that can feasibly be measured and balanced. Choosing which attributes to represent is therefore unavoidable. Evidence about which characteristics citizens value informs which attributes should be included in a quota schema. Understanding how citizens combine those characteristics into judgments of representation then provides a basis for interpreting the guarantees of different selection methods and auditing alternative panel-selection algorithms. 

Throughout the paper, we focus on demographic characteristics such as gender and age, together with easily elicited attitudinal characteristics such as political alignment and views on the issue under consideration. These are straightforward to collect from participants and are commonly used, or could readily be incorporated, in the design of citizens' assemblies.

\subsection{Empirical Approach and Main Findings}

As a first step toward answering these empirical  questions, we conducted two UK conjoint surveys covering three tasks. To measure how citizens evaluate potential assembly members, respondents repeatedly chose which of two hypothetical candidates for a citizens' assembly would better represent them. Each candidate was described by seven common attributes: gender, age group, area type,
ethnicity, disability status, education, and UK region. Many of these attributes have been systematically used to recruit participants for previous citizens' assemblies
\citep{bcassembly2004,climateassemblyuk2020}. In addition to these common attributes, we varied party affiliation and climate concern to examine political and issue-specific dimensions of representation.

Survey~1 asked 1,151 respondents to evaluate profiles for a general-purpose assembly in which party affiliation was displayed and a climate-focused assembly in which climate concern was displayed. Survey~2 asked 576 respondents to evaluate climate-focused profiles displaying both attributes. In each task, respondents first reported their own attributes and then made repeated pairwise choices between randomly generated candidate profiles in a randomized conjoint experiment \citep{mcfadden1972conditional,louviere1983design,hainmueller2014causal,hainmueller2015immigration}. These choices form the basis of our primary analyses.

\paragraph{Attribute elicitation.}
We estimate average marginal component effects (AMCEs), following standard practice in conjoint analysis \citep{hainmueller2014causal,bansak2023amce} and recent recommendations for interpreting subgroup analyses \citep{delacuesta2022external,leeper2020measuring}.
The main takeaway is that the importance of different characteristics is far from uniform. Although demographic characteristics influence perceived representation, respondents place substantially greater weight on attitudinal characteristics such as political alignment and, in issue-specific settings, views on the issue under consideration. In the general-purpose assembly, political alignment is by far the strongest determinant of perceived representation. In the climate-focused assembly, climate concern becomes the dominant characteristic when political alignment is not displayed. However, when both characteristics are shown, political alignment again emerges as the primary driver of representation judgments. Climate concern continues to exert a stronger influence than any demographic characteristic.
The subgroup estimates show recurring patterns in the observed data.
Across all three tasks, the estimated age-mismatch penalty is largest among
respondents aged 16--29, and the estimated gender-mismatch penalty is larger
among female than male respondents. These are descriptive comparisons of
estimated effects, not formal tests of between-group differences.
Value-specific estimates also show directional and distance-sensitive patterns
for ordered attributes.

Survey~1 invited general feedback on the survey. Some respondents used that broader
feedback field to question demographic resemblance and emphasize shared views,
ways of reasoning, and deliberative qualities. Separately, Survey~2 also included a targeted open-ended question asking which
additional attributes might help measure representation. We treat the Survey~1 comments only as illustrative
feedback and analyze the Survey~2
attribute suggestions.

\paragraph{Omission stress test.}
Suppose organizers decide not to include a characteristic that citizens care about in the quota schema. Can that omitted characteristic nevertheless be represented indirectly through correlated demographic or political attributes, or must it be represented explicitly?
Quota-based sortition methods provide fairness and robustness guarantees within
a declared quota schema
\citep{flanigan2021fair,flanigan2024transparent,assos2025alternates,
gambhir2025dropout}. However, classical stratification suggests that correlated quota attributes may offer some indirect protection for omitted characteristics when they are unevenly distributed across the observed strata~\citep{benade2019stratification}. This suggests that correlated
quota attributes may provide some indirect protection. Using the 1,151 Survey~1 respondents as
the pool, we implement the LEXIMIN procedure of \citet{flanigan2021fair}. We compare uniform sampling with demographic quotas, demographic-plus-party quotas, and quotas that also constrain climate concern. Adding party affiliation to demographic quotas provides little improvement in representing the omitted climate-concern distribution. In contrast, explicitly constraining climate concern substantially reduces representation error.   The main takeaway is that correlated demographic and political characteristics cannot always substitute for omitted dimensions of representation. Consequently, identifying which characteristics citizens consider important is a prerequisite for designing effective quota schemes.

\paragraph{Predicting representation
choices.}  Lastly, we ask how well respondent and candidate attributes predict representation choices. This analysis serves two purposes. First, it provides a practical tool for auditing panel-selection methods by estimating whether a proposed panel contains members whom citizens would regard as good representatives of themselves. Second, it tests whether a simple shared metric can adequately describe representation judgments, providing empirical evidence for a central behavioral assumption underlying recent computational social choice approaches to representative selection \citep{ebadian2022sortition,ebadian2025boosting,kalayci2024proportional}. 
We compare transparent matching heuristics with two modeling approaches. Metric models represent choice as proximity in a shared attribute space. Utility models describe choice through the relative utility of competing profiles. We use a covariate-extended Bradley--Terry formulation, in which candidate utilities vary with respondent attributes \citep{bradley1952rank,dittrich1998subject}; closely related Bradley--Terry likelihoods are also widely used in recent human-preference learning \citep{rafailov2023direct,hong2024adaptive}.
We evaluate the models on held-out respondents. 
Both  the learned metric and the
conditional Bradley--Terry model  outperform single-cue, unweighted, and fixed-priority rules.  Although the respondent-conditioned Bradley--Terry model achieves slightly higher predictive accuracy than the metric model, the improvement is modest despite its substantially greater complexity. 

\subsection{Related Work}
\label{sec:related-work}

\paragraph{Political representation.}
Studies of political candidates, representatives, and representative claims
find that shared party affiliation can be central to perceived representation
 and can matter more than demographic resemblance
\citep{arnesen2019inferences,blumenau2025multidimensional,
vik2025unelected}. This motivates testing political and topic-specific alignment
alongside demographic attributes in citizens' assemblies. Conjoint experiments
have likewise decomposed multidimensional candidate choice, examining
interactions and trade-offs among ideological, valence, personal, and
social-role attributes
\citep{franchino2015voting,horiuchi2020personal,teele2018ties}. Whereas these
studies elicit electoral preferences, we ask which attributes lead respondents
to view prospective assembly members as representing them.  

\paragraph{Representation and legitimacy in mini-publics.}
The mini-public literature has primarily studied how the wider public evaluates
an assembly as a whole. Conjoint studies vary institutional
rather than participant attributes, showing how features such as inclusiveness,
deliberative format, transparency, decision authority, and autonomy shape
support for participatory processes and mini-publics
\citep{christensen2020participatory,goldberg2025empowered}. Perceiving assembly participants as ``people like me''
is associated with greater legitimacy and willingness to accept the
assembly's outcome~\citep{pow2020likeme}. Experiments that vary information
about panel composition also show that demographic and attitudinal  attributes imbalances
can reduce legitimacy, although composition effects are not uniform across
studies or respondents~\citep{germann2025representativeness,
paulis2024fair}. This work establishes that similarity and aggregate
representativeness can matter, but it does not isolate which attributes lead a
person to view a prospective citizen representative as representing them. We
examine that question directly using randomized candidate profiles that
distinguish demographic, political, and topic-specific attributes. 

\section{Representation and Experimental Design}
\label{sec:conjoint-design}

We use two randomized conjoint surveys to study how people judge whether a potential assembly member represents them. This section first defines the representation space on which panel-selection procedures operate, then describes the assembly contexts, samples, randomized profiles, and survey tasks.

\paragraph{What is being represented?}

Let $U$ be the population that an assembly is intended to resemble, let $S \subseteq U$ be a selected panel of fixed size $|S|=k$, and let $A$ be a selected set of attributes. Each attribute $a\in A$ takes values for a  finite  set $L_a$ and a person's attribute profile is a vector $\vec x_i = (x_{ia})_{a \in A}$ in the representation space $\X = \prod_{a\in A} L_a$. 

Quota-based selection, the predominant approach in
citizens' assembly practice, specifies lower and upper bounds
$\underline q_{a\ell}$ and $\overline q_{a\ell}$ for each value $\ell\in L_a$.
Writing $n_{a\ell}(S)=\sum_{i\in S}1\{x_{ia}=\ell\}$, a panel satisfies the
marginal quota constraints when
$\underline q_{a\ell}\leq n_{a\ell}(S)\leq\overline q_{a\ell}$ for every $a$
and $\ell$ \citep{flanigan2021fair,flanigan2024transparent}.
These methods require organizers to specify the attributes on which the panel should be stratified before the selection algorithm can be applied. However, it is very unclear  how the attribute set itself should be chosen. Our work addresses this missing step by studying \emph{perceived representation}: given two candidate profiles, which person does an individual believe would better represent them
as someone who shares your background, experiences, and concerns? By examining how individuals perceive representation, we aim to inform the selection of relevant attributes and the formulation of representation objectives for future panel-design methods.

\subsection{Survey: Topics and Samples}

We conducted two conjoint surveys examining general-purpose and climate-focused citizens' assemblies. After filtering, Survey~1 retained 1,151 respondents, each of whom evaluated candidates for both a general-purpose citizens' assembly and a citizens' assembly on climate change. Candidate profiles in both tasks displayed seven shared demographic and geographic attributes---gender, age group, area type, ethnicity, disability status, education, and UK region. The general-purpose task additionally displayed political alignment, whereas the climate-focused task displayed climate concern. Respondents completed five pairwise comparisons per topic, yielding 5,755 direct choices per topic.\footnote{Survey~1 also included two ranking tasks per topic. Because responses to the ranking tasks were considerably noisier than the direct pairwise choices, we do not use them in the analyses reported here.} Survey~2 retained 576 respondents, who completed ten pairwise comparisons for a climate-focused assembly in which both political alignment and climate concern appeared alongside the seven shared attributes, yielding 5,760 direct choices.

Survey~1
was fielded across four UK waves and Survey~2 across two. In both surveys, we
excluded completion times below the 2nd or above the 98th percentile of the
otherwise-eligible completion-time distribution.
Further details on recruitment, exclusions, sample composition, and data
cleaning appear in the supplementary material.  
Before the survey tasks, the landing page provided a short research-team-produced explainer video and a written introduction to citizens' assemblies. A dedicated instruction screen then specified the remit of each assembly before its comparison tasks; the full participant-facing context and delivery sequence are reported in the supplementary material. The surveys differed in the attributes shown to respondents, and candidate profiles were randomized within each task. 

\subsection{Randomized Candidate Profiles and Survey Tasks }

Across the two surveys, respondents reported their own
attributes and made pairwise comparisons between candidates. Their final
free-text items differed: Survey~1 invited general feedback on the survey,
whereas Survey~2 specifically asked whether other attributes would help measure
representation. In the self-attribute-reporting task, respondents were shown
each attribute one at a time and asked to select the category that best
described them. In the pairwise-comparison task, candidates were described by
the attributes displayed for that assembly context. Their profiles were
constructed by randomly assigning a category value to each displayed attribute
according to the procedure described below. 

In Survey~1, and for the seven shared
attributes in Survey~2, candidate attributes are randomized relative to the
respondent's profile $\vec p$. For each attribute, the candidate matches the
respondent with probability $0.5$; conditional on a mismatch, each of the
remaining values is equally likely:
\[
\Pr(c_a=p_a)=\tfrac12,\qquad
\Pr(c_a=\ell)=\tfrac{1}{2(|L_a|-1)}\quad(\ell\ne p_a).
\]
This design gives respondents with common and rare attribute values the same
probability of seeing a match. In Survey~1, party affiliation and level of
climate concern are sampled independently because they are not displayed
together in the same candidate profile. Candidate position and the displayed
order of attributes are also randomized.

In Survey~2, party alignment and climate concern appear together in each candidate profile. 
Prior research documents an  association between political affiliation and climate attitudes~\cite{cruz2017relationships, gregersen2020political}. Sampling their values independently would ignore this association and could produce combinations that are rare in the population. We therefore draw party--concern pairs from a smoothed empirical joint distribution estimated from Survey~1, subject to the match--mismatch randomization described below.
Let $C_{rs}$ denote
the party-by-concern count matrix from Survey~1. We smooth the empirical
distribution slightly to ensure that every combination can be sampled:
\[
 \widetilde p_{rs}=(1-\lambda)\frac{C_{rs}}{\sum_{uv}C_{uv}}
 +\lambda\frac{1}{8\cdot4},\qquad \lambda=0.05.
\]
For each candidate, the party-match and concern-match indicators are first
drawn independently, each with probability one-half. 
These two draws determine which
party--concern pairs are eligible. For example, if party must match and concern must differ, we retain only pairs with the respondent's party and a different concern level. We renormalize $\widetilde p$ over the eligible pairs and sample one. Thus, each of the four match--mismatch combinations occurs with probability one-quarter, while eligible pairs remain weighted by their probabilities in the smoothed empirical distribution. The count matrix and complete conditional sampling procedure are
provided in the supplementary material.

\section{Which Attributes Shape Perceived Representation?}
\label{sec:amce}

Here, we examine which attributes people use when deciding whether one prospective assembly member would represent them better than another. We estimate the effect of mismatches on individual attributes and compare the relative importance of demographic, political, and issue-specific dimensions of representation.

\begin{figure*}[t]
  \centering
  \includegraphics[width=0.98\textwidth]{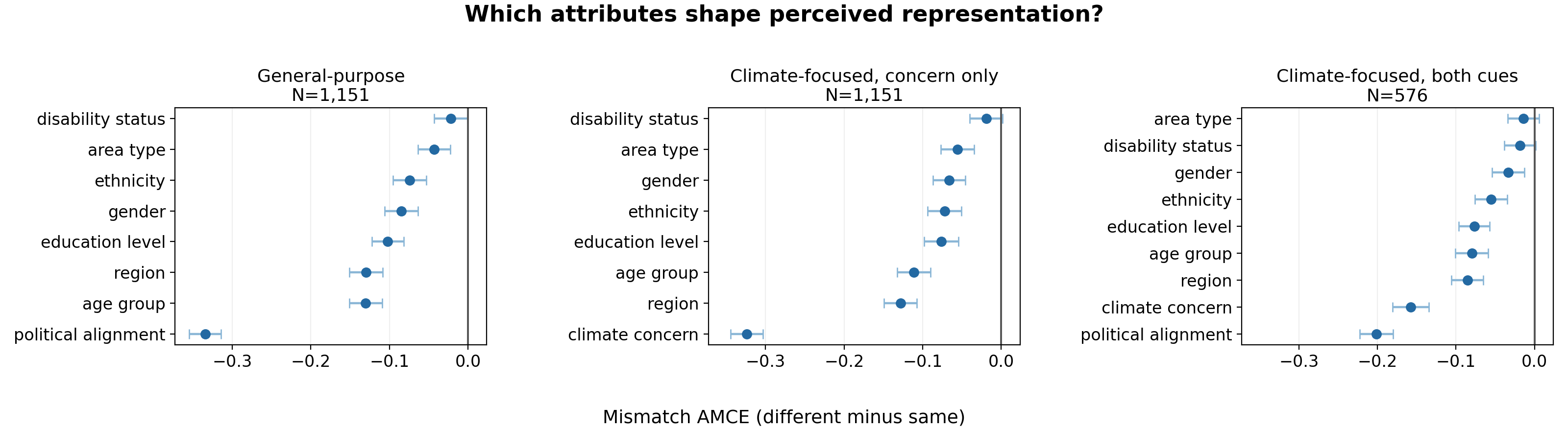}
  \caption{Mismatch AMCEs from randomized direct choices. Points show the mean
  change in selection probability when an attribute differs from, rather than
  matches, the respondent; bars are 95\% confidence intervals.}
  \Description{Three forest plots of mismatch effects for general-purpose,
  climate-focused, and both-cues climate assembly tasks. Political alignment
  and climate concern have the largest negative effects.}
  \label{fig:amce-main}
\end{figure*}

\subsection{The Relative Importance of Candidate Attributes}

\paragraph{Estimating mismatch penalties.}

We measure an attribute's importance by the average marginal component effect (AMCE) of switching it from a match to a mismatch, while averaging over the randomized distribution of the other profile features \citep{hainmueller2014causal,leeper2020measuring,bansak2023amce}.

For respondent $i$ and candidate $j$, let $\vec p_i=(p_{ia})_{a\in A}$ and $\vec c_j=(c_{ja})_{a\in A}$ denote their respective profiles, and define the mismatch indicator $M_{ija}=1\{c_{ja}\ne p_{ia}\}$. Let $\mathbf M_{ij,-a}=(M_{ijb})_{b\in A\setminus\{a\}}$ collect the remaining mismatch indicators. For $t\in\{0,1\}$ and a realization $\mathbf u$ of this mismatch indicator vector, let $Y_{ij}(t,\mathbf u)$ be the potential choice outcome under those mismatch assignments, averaged over the candidate values generated conditional on them. The mismatch AMCE for attribute $a$ is defined as
\[
 \tau_a=\mathbb E_{i,j,\mathbf M_{ij,-a}}\!\left[
 Y_{ij}(1,\mathbf M_{ij,-a})-Y_{ij}(0,\mathbf M_{ij,-a})\right].
\]
The same realization of $\mathbf M_{ij,-a}$ appears in both terms: the counterfactual contrast changes only the focal mismatch state and then averages over the randomized states of the other attributes. Respondents need not see these two counterfactual profiles as a literal pair. Because mismatch events are independently randomized across attributes, this contrast identifies $\tau_a$. In the both-cues experiment, the party and climate-concern values are sampled jointly, but their mismatch indicators remain independent; the raw values are averaged over their conditional sampling distribution above.

To weight respondents equally, let $\overline Y_i(M_a=t)$ be respondent $i$'s
mean choice outcome in state $t$, and let $\mathcal I_a$ contain those who
encounter both states. We estimate
\[
 \widehat\tau_a=\frac{1}{|\mathcal I_a|}
 \sum_{i\in\mathcal I_a}\left[
 \overline Y_i(M_a=1)-\overline Y_i(M_a=0)\right].
\]
We compute 95\% confidence intervals; details are provided in the supplementary
material.
More negative values indicate larger penalties for mismatch and, therefore,
greater importance for perceived representation.

\paragraph{Variation in estimated attribute importance.}

Figure~\ref{fig:amce-main} shows that candidate--respondent mismatch
generally reduces selection probability, but the magnitude varies markedly by
attribute. Political alignment has the largest estimated penalty in the
general-purpose task, whereas climate concern has the largest penalty in the
concern-only Climate task. When both cues are displayed, they remain the two
largest, with a greater estimated penalty for party mismatch. Age and region
generally form a second tier, while area type and disability status have
smaller estimated effects. Because each effect is identified within a
randomized task, the within-task contrasts are causal; comparisons across task
environments remain descriptive. Overall, political and issue-specific
characteristics carry more weight than most demographic characteristics, but
their relative importance differs across the observed decision contexts.

The average effects also conceal descriptive variation
across respondent groups and across particular mismatches. We therefore
examine the clearest recurring patterns before turning to predictive models,
while treating them as exploratory rather than confirmatory.

\subsection{Exploratory Heterogeneity in Representation Judgments}
\label{sec:amce-heterogeneity}

\begin{figure*}[t]
  \centering
  \includegraphics[width=0.96\textwidth,keepaspectratio]
    {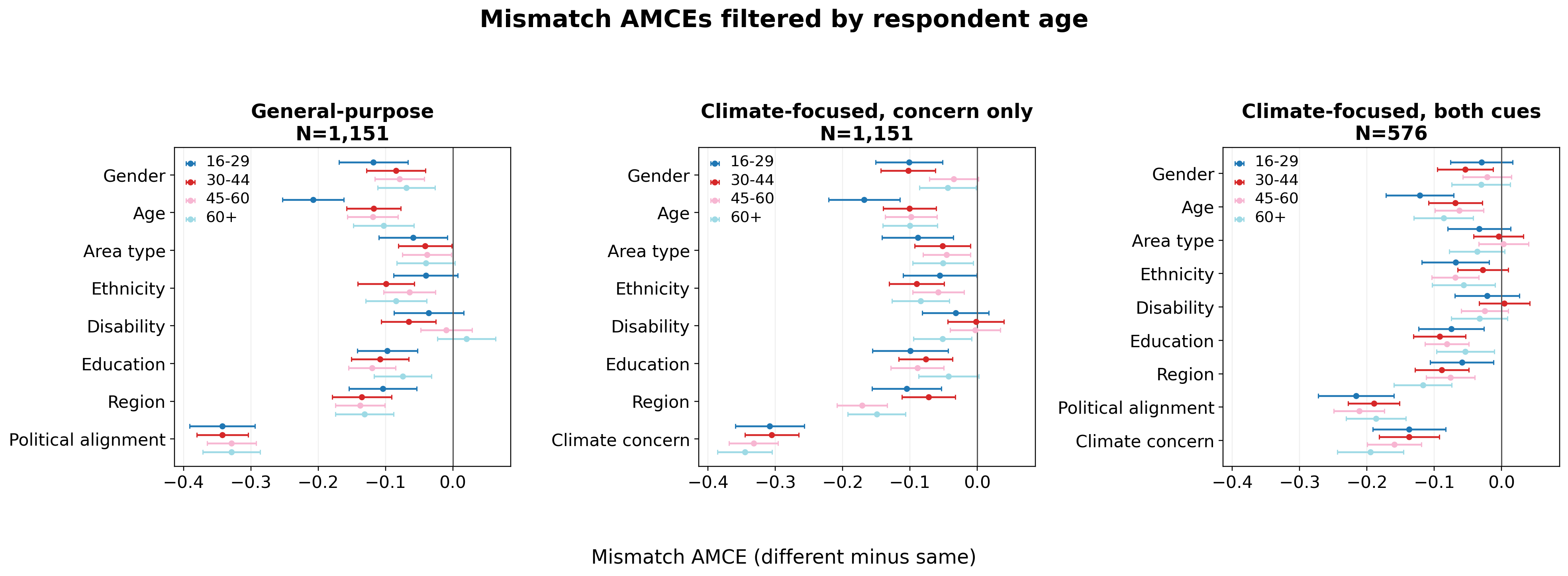}\\[-2mm]
  \includegraphics[width=0.96\textwidth,keepaspectratio]
    {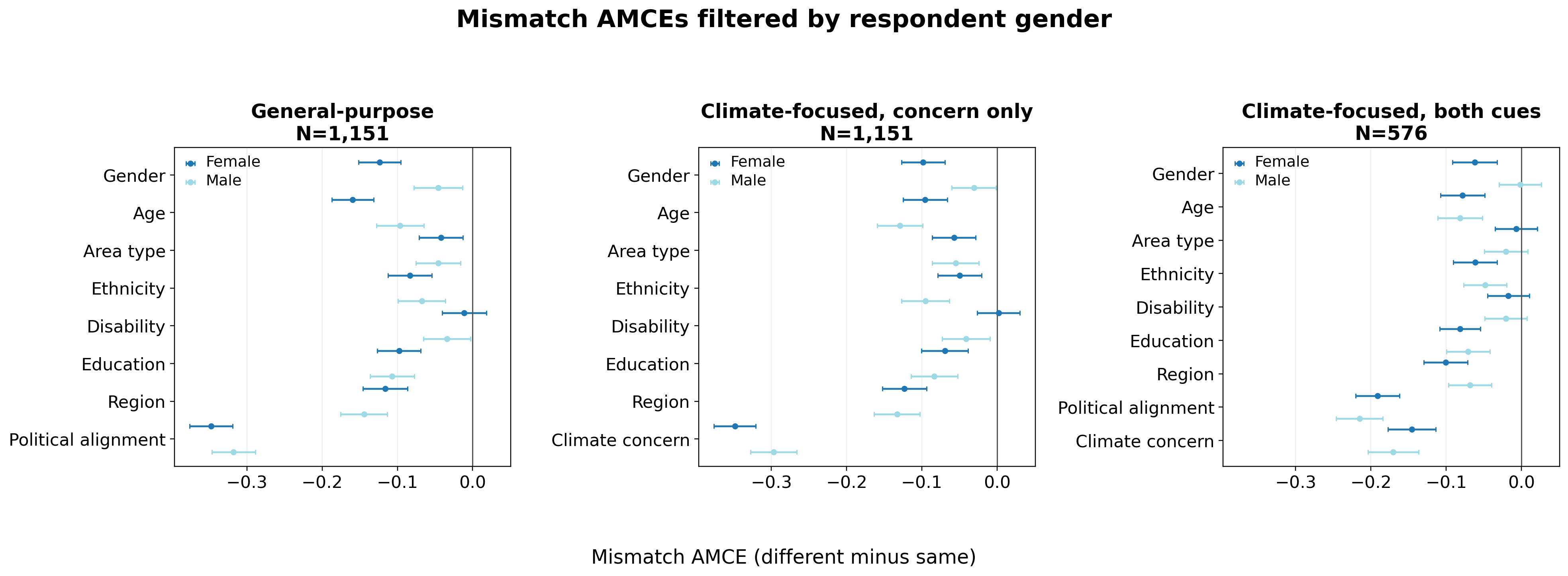}
  \caption{Mismatch AMCEs estimated by respondent subgroup across the
  three task environments. The upper plot separates respondents by age and the
  lower plot by gender. Each panel reports candidate-mismatch effects for every
  displayed attribute; points show estimates and bars show pointwise 95\%
  confidence intervals.}
  \Description{The upper three forest plots compare mismatch AMCEs
  across four respondent age groups; the lower three compare female and male
  respondents. The comparisons cover the general-purpose, concern-only
  climate, and both-cues climate tasks.}
  \label{fig:app-filtered-age}
  \label{fig:app-filtered-gender}
\end{figure*}

We first re-estimate every mismatch AMCE separately within respondent
age and gender groups. Figure~\ref{fig:app-filtered-age} shows both filtered
analyses across all three tasks. These analyses are exploratory: intervals are
pointwise, we do not adjust for the number of comparisons, and we do not
interpret visual differences as formal interaction tests. The supplementary
material gives the estimator, eligibility thresholds, and exact focal
estimates in Table~\ref{tab:appendix-age-gender-amce}.

Two patterns recur in the subgroup point estimates. The age-mismatch
penalty is largest among respondents aged 16--29 in each task: $-.208$ in the
General task, $-.168$ in the concern-only Climate task, and $-.121$ in the
both-cues task. The estimated gender-mismatch penalty is also larger among
female than male respondents in each task. In the General task, for example,
the estimates are $-.124$ and $-.046$, respectively. These recurring
differences suggest that the perceived importance of a candidate attribute
may depend partly on the respondent's own profile, but the pointwise intervals
do not constitute formal tests of between-group differences.

\begin{figure*}[t]
  \centering
  \includegraphics[width=0.46\textwidth,keepaspectratio]
    {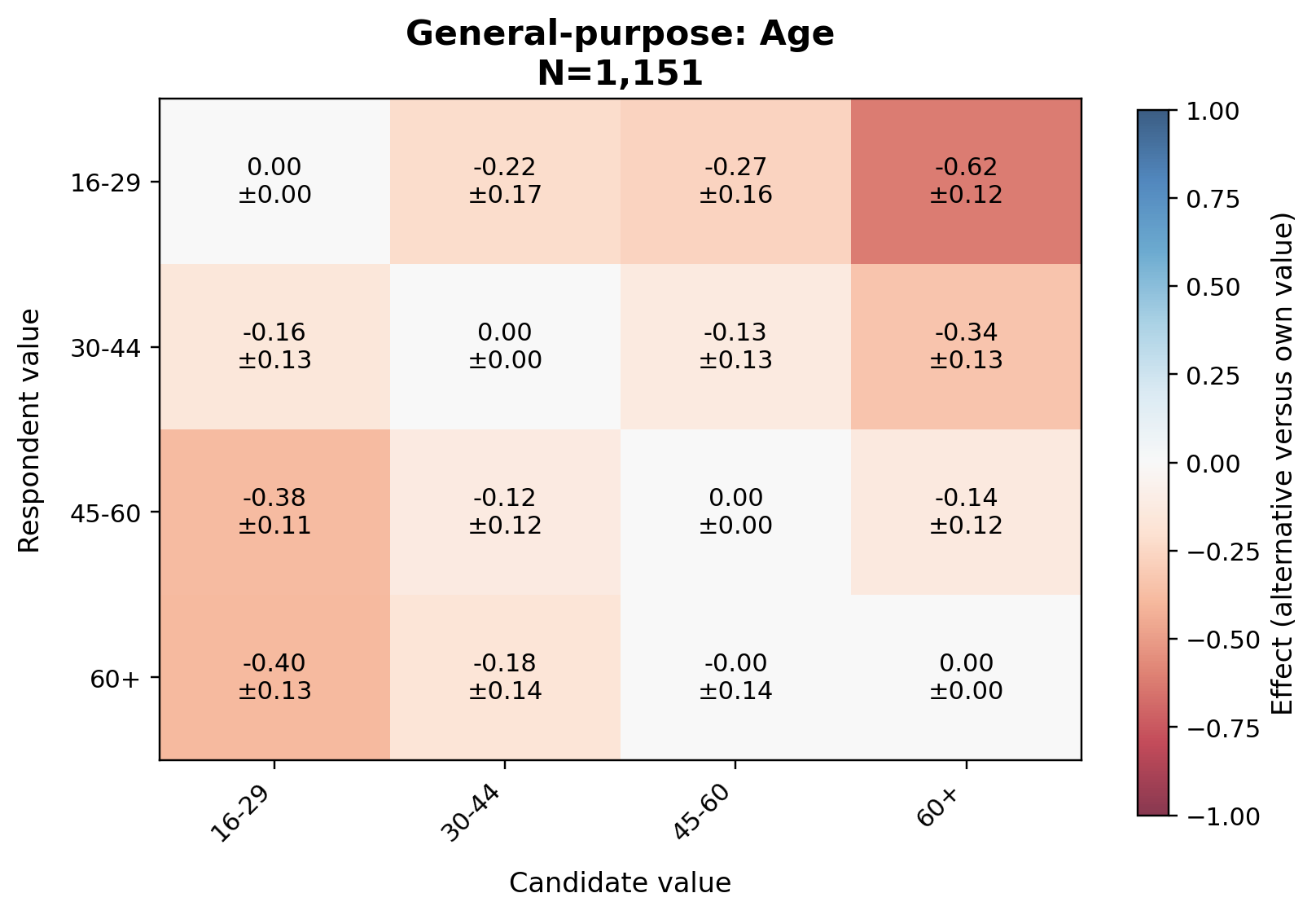}\hfill
  \includegraphics[width=0.46\textwidth,keepaspectratio]
    {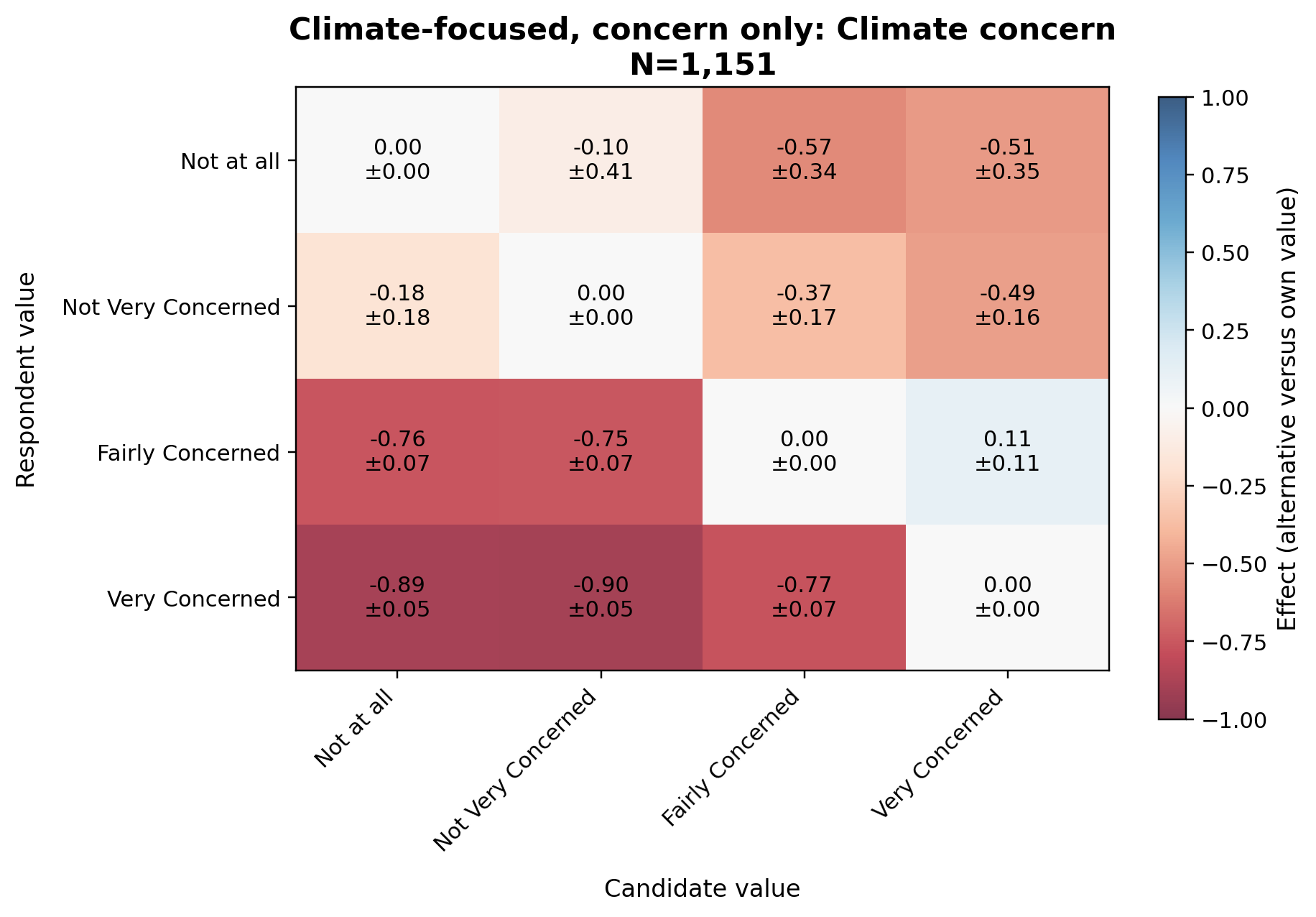}
  \caption{Group-conditioned own-versus-alternative effects for age in
  the general-purpose task (left) and climate concern in the Survey~1
  concern-only climate task (right). Text in each cell gives the
  respondent-averaged effect and pointwise 95\% interval half-width. Both plots
  use the common effect scale shown in their color bars.}
  \Description{Two heatmaps compare respondent values in rows with
  candidate values in columns. More negative cells indicate a stronger
  preference for the respondent's own age group or climate-concern level.}
  \label{fig:app-heatmap-general-a}
  \label{fig:app-heatmap-climate-concern}
\end{figure*}

We next disaggregate the binary mismatch indicator for two ordered
attributes. Each cell in Figure~\ref{fig:app-heatmap-general-a} compares a
candidate carrying the column value with one matching the respondent's row
value and reports $2\Pr(\text{column-value candidate chosen})-1$. Negative
values therefore favor the respondent's own value. Formal definitions and
support thresholds appear in the supplementary material.

The age point estimates display an ordinal gradient: adjacent groups
are generally more acceptable than groups farther apart, and the youngest
respondents exhibit the clearest decline as candidate age increases. The
climate-concern estimates are also ordered but directionally asymmetric. Very
concerned respondents strongly favor their own value over lower-concern
candidates, whereas fairly concerned respondents do not comparably favor their
own value over more-concerned candidates. The ordinal patterns motivate using
graded gaps rather than binary mismatches in the weighted-$\ell_1$ model. The
directional pattern also identifies a feature that a shared symmetric distance
cannot represent. Political-alignment estimates show further pairing-specific
asymmetry and appear in
Figure~\ref{fig:app-heatmap-general-politics} in the supplementary material.

\subsection{Open-Ended Attribute Elicitation and Survey Feedback}
\label{sec:open-ended-elicitation}

The conjoint tasks necessarily restrict representation to attributes chosen in
advance. At the end of each survey, we therefore included an optional
open-ended question. In Survey~1, we invited respondents to provide general
feedback:
\begin{quote}\small
``Have questions about the survey or our research? We'd love to hear from you.
Send us your comments, inquiries, or feedback.''
\end{quote}
Two responses raised particular concerns about using profile attributes to
measure representation. The first argued that demographic matching could
encourage stereotyping while missing shared views:
\begin{quote}\small
``I couldn't care less about my representative's background, just whether they
agree with my views. You end up stereotyping to try and get that result, which
is grim.''
\end{quote}
The second argued that static attitudes and profile characteristics do not
capture how a representative reasons and makes decisions:
\begin{quote}\small
``So if I am choosing who may represent me, I want to understand more than just
set attitudes and static opinions; how and why they think and decide is
relevant.''
\end{quote}

In Survey~2, we asked a more targeted question about attributes omitted from
the displayed profiles:
\begin{quote}\small
``Are there any other attributes that would help measure how well someone
represents you in a citizens' assembly, either on climate change or in general?
Please share any suggestions.''
\end{quote}
Among the 576 Survey~2 respondents retained after screening, 261 (45.3\%)
provided a non-empty answer, while 64 explicitly stated that no additional
attribute was needed. After treating answers that only repeated the displayed
political-alignment or climate-concern cues separately, 179 respondents (31.1\%
of the full sample) named at least one additional theme. Because a response
could mention several themes, we used multi-label coding.

The most frequent suggestions concerned deliberative qualities such as
listening, openness, and constructive participation (27 respondents), followed
by socioeconomic position (22), occupation or work experience (20), family or
household status (19), local or community connection (19), general capability
or education (19), and climate knowledge or evidence orientation (17). These
nominations extend beyond demographic resemblance to how representatives
deliberate and to the circumstances, experience, and knowledge they bring to
the assembly. Seven Survey~2 respondents also de-emphasized one or more
displayed demographic attributes. One distinguished personal resemblance from
the group-level purpose of inclusion:
\begin{quote}\small
``I am not sure that ethnicity, gender or disability status should be
considered as an attribute when representing me apart from ensuring that there
is representation from these groups.''
\end{quote}
Together, the comments distinguish group-level inclusion, which demographic
quotas can safeguard, from individual judgments of representation, which may
also depend on substantive agreement and deliberative capacity. Because both
free-text items were optional and the comments were self-selected, they identify
considerations for future study rather than their prevalence or a ready-made
set of additional quotas. Full coding details appear in the supplementary
material.

\section{Can Topic-Specific Representation Be Achieved Indirectly?}

\label{sec:sampling}

A central design question in quota-based selection is whether every relevant attribute must be explicitly included in the quota schema. Collecting additional attributes increases logistical and privacy costs, and some attributes may be difficult elicit. A natural question is therefore whether an omitted attribute can be represented indirectly through its correlation with the attributes that are already included. If so, the included attributes may serve as adequate proxies; if not, a panel can satisfy every stated quota while remaining systematically unbalanced along an important omitted dimension.

\subsection{Party Quotas as a Proxy for Climate Concern}

To investigate whether omitted attributes can be adequately represented by correlated included attributes, we consider climate concern as a case study. This provides a relatively favorable setting for the proxy hypothesis, in Survey~1, party affiliation and climate concern exhibit a moderate association (Cramér's V=0.261; see the supplementary material for the computation). We therefore use the Survey~1 pool to test whether adding party to a demographic quota schema preserves the distribution of climate concern more closely than demographic quotas alone or uniform random sampling without quota constraints. Panels that also constrain climate concern provide a direct-control benchmark.

\paragraph{Quota-constrained panel selection.}
We draw 250 panels of size 40 from the 1,151 respondents in
the Survey~1 pool.
Let $U$ denote this observed pool for the experiment; it is the reference
distribution, not the UK population.
We compare uniform sampling with panels constrained on (i) seven demographics,
(ii) demographics and party, or (iii) demographics, party, and concern. Pool
shares are multiplied by 40 and rounded down; any remaining seats are assigned
to the categories with the largest fractional remainders. Each resulting target
has lower and upper bounds one seat below and above it. For each schema, we apply
the released LEXIMIN implementation of~\citep{flanigan2021fair}. LEXIMIN computes a distribution over
quota-feasible panels that lexicographically maximizes individual inclusion
probabilities, beginning with the least likely person. We sample from the
returned distribution using a fixed seed. Implementation details are provided
in the supplementary material. 

Let $U$
be the observed pool, $S\subseteq U$ a selected panel, and $n_\ell(V)$ the
number of members of $V$ at concern level $\ell$. The omitted-attribute error is
\[
 \operatorname{TV}(S,U)=\tfrac12\sum_{\ell}
 \left|\frac{n_{\ell}(S)}{|S|}-\frac{n_{\ell}(U)}{|U|}\right|.
\] TV is zero
when the panel and pool concern distributions match. Since pool
shares rarely correspond to integer counts in a 40-person panel, the uniformly
random panels provide a natural finite-sample benchmark.

\paragraph{Results.}

\begin{figure}[t]
  \centering
  \includegraphics[width=\columnwidth]{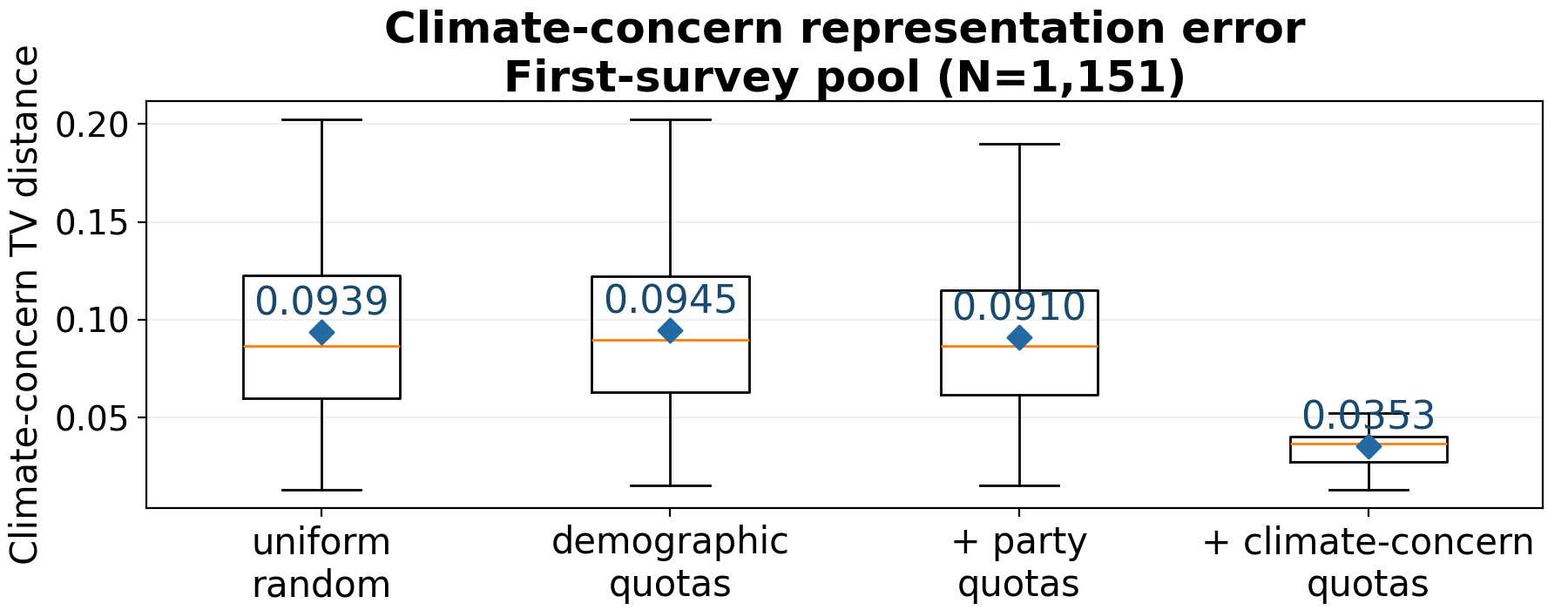}
  \caption{Climate-concern error for 250 panels of size 40 from the observed
  Survey~1 respondent pool. Boxes show
  pool-relative TV distance; diamonds and labels show means. Quota panels are draws from the
  released LEXIMIN procedure with bounds one seat around proportionally rounded
  marginal targets.}
  \Description{Boxplots show similar
  climate-concern errors for random, demographic-quota, and
  demographic-plus-party panels, and lower errors when climate concern is
  directly constrained.}
  \label{fig:sampling-tv}
  \label{main-fig:sampling-tv}
\end{figure}

Mean TV is 0.0939
under uniform sampling, 0.0945 with demographic quotas, and 0.0910 when party
quotas are added. Despite the observed association, adding party quotas reduces
mean TV by only 0.0035 relative to demographic quotas. Direct concern quotas
reduce mean TV to 0.0353.
We check whether these findings depend on the sampling procedure, panel size,
or permitted deviation from quota targets. Across the tested specifications,
directly constraining climate concern consistently improves concern balance
relative to uniform random sampling. Demographic quotas, with or without party,
provide no consistent improvement. Full feasibility and error results are
reported in the supplementary material.

The key result is that demographic and political quotas cannot reliably substitute for a relevant topic-specific attribute. Although political alignment is correlated with climate concern, adding party quotas yields only a marginal improvement over uniform random sampling and no consistent improvement across our robustness analyses. By contrast, directly stratifying on climate concern substantially reduces imbalance. This suggests that correlation alone is insufficient to ensure representation of an omitted attribute.

\section{Predicting Representation Choices}
\label{sec:structural}

The AMCE analysis establishes which displayed attributes change perceived
representation on average. It does not, by itself, describe how those
attributes are combined when a respondent chooses between two profiles.
We therefore move from marginal effects to joint prediction, asking how well
alternative models predict an individual's choice between two profiles.
Performance on held-out respondents provides a common criterion for assessing
how much structure is needed to summarize these choices. The models range from
simple heuristics to a flexible, respondent-conditioned choice model.

This analysis serves two purposes.
First, it tests a structural assumption made by recent metric-based theories of representation~\cite{ebadian2022sortition, ebadian2025boosting, kalayci2024proportional, kalayci2026temporal}. These works offer techniques for achieving intersectional representation
(representation across combinations of attributes, rather than each
attribute in isolation) in settings where guaranteeing every combination
explicitly is infeasible. Their guarantees rest on the premise that a
single shared metric can stand in for how people judge similarity; our
prediction task tests that premise directly. Second, a model that
predicts choices well can be useful independently of any particular selection
algorithm. For an observed respondent profile, it estimates which of two
candidate profiles is more likely to be viewed as the better representative.
Aggregated across respondent profiles, such estimates could complement
quota-based audits by identifying groups for whom the available candidates
are systematically less representative.

\subsection{Predictive
Models}

\paragraph{Heuristic models.}
We begin with three simple, interpretable baselines. \emph{Attribute match count} selects
the candidate with more exact matches across the displayed attributes.
\emph{Most-important cue} predicts choices using only the attribute with the
largest absolute mismatch AMCE estimated from the training data;
this is party in every General and both-cues training split and climate
concern in every concern-only Climate split.
The \emph{AMCE-ordered match tree} instead inspects all attributes
in decreasing training-AMCE magnitude and stops at the first one on which
exactly one candidate matches the respondent. Unresolved comparisons receive a
symmetric tie. Complete definitions and implementation details are given in the
supplementary material.

\paragraph{Weighted $\ell_1$ metric.}
The fourth model treats perceived representation as proximity under
\[
  d_{\vec w}(\vec p,\vec c)
  =
  \sum_{h\in\mathcal A}w_h\delta_h(p_h,c_h).
\]

Here $\delta_h$ measures the per-attribute gap---how far
apart the candidate and respondent are on attribute $h$---using an ordinal gap for age,
education, and climate concern (so a two-step difference counts more than
a one-step difference) and a simple match--mismatch distance for every
other attribute. The weight $w_h \geq 0$, learned from the data, controls
how much that attribute's gap matters relative to the others. Prediction then follows directly from which
candidate is closer,
with ties scored as one-half. We fit the weights with a margin-one hinge
objective and $\ell_2$ regularization; that is, we push the chosen
candidate to be closer than the rejected one by at least a fixed margin,
penalizing violations linearly. Full objective and implementation
details appear in the supplementary predictive-model appendix. 

\paragraph{Conditional BT.}
Having fit a single common ruler for all respondents, we next ask whether
that restriction costs predictive power. The fifth model relaxes it using
a covariate extension of Bradley--Terry \citep{bradley1952rank}, allowing the
importance of a candidate's attributes to vary with the respondent's own
profile rather than being fixed across everyone. Such models allow
paired-comparison utilities to depend jointly on subject and candidate
attributes \citep{dittrich1998subject}. Let $\vec x_{c_1}$ and
$\vec x_{c_2}$ encode the candidate profiles, $\vec z_p$ encode the
respondent, and $\vec d=\vec x_{c_1}-\vec x_{c_2}$. Then
\[
  \Pr(\vec c_1\succ\vec c_2\mid\vec p)
  =
  \sigma\!\left(
    \vec\beta^\top\vec d+\vec z_p^\top W\vec d
  \right),
\]
where $\vec\beta$ captures average candidate-value preferences, shared
across all respondents, and $W$ lets those preferences shift with the
respondent's observed profile: a nonzero entry means that how much a given
candidate-attribute gap matters depends on the respondent's own value for
that attribute. Our
implementation restricts $W$ to same-attribute blocks
(e.g., respondent age by candidate age), omitting cross-attribute
interactions to keep the model tractable and interpretable. It is fitted by L2-regularized logistic regression without participant identifiers and has 268, 224, and 280
coefficients in the General, concern-only Climate, and both-cues tasks,
respectively, versus 8, 8, and 9 weighted-$\ell_1$ coefficients. Complete encoding and
estimation details appear in the supplementary predictive-model appendix.

\FloatBarrier
\begin{table*}[t]
  \centering
  \scriptsize
  \setlength{\tabcolsep}{4pt}
  \caption{Predictive accuracy across all three task environments (mean $\pm$ s.d.\
  over 20 shared 85/15 participant-level splits).
  ``\# Parameters'' counts fitted choice-rule coefficients or, for the AMCE
  tree, training-derived attribute scores.}
  
  \label{tab:lp-results}
  \begin{tabular}{@{}llrrr@{}}
    \toprule
    Task & Model & \# Parameters & Train acc. & Test acc. \\
    \midrule
    General-purpose
      & Attribute match count & 0 & $.669\pm.001$ & $.666\pm.008$ \\
      & Party cue only & 0 & $.633\pm.002$ & $.634\pm.009$ \\
      & AMCE-ordered tree & 8 & $.705\pm.003$ & $.697\pm.017$ \\
      & Weighted $\ell_1$ & 8 & $.732\pm.003$ & $.725\pm.015$ \\
      & Conditional BT & 268 & $.772\pm.003$ & $\mathbf{.738\pm.015}$ \\
    \midrule
    Climate, concern only
      & Attribute match count & 0 & $.648\pm.003$ & $.652\pm.017$ \\
      & Climate cue only & 0 & $.644\pm.003$ & $.646\pm.015$ \\
      & AMCE-ordered tree & 8 & $.711\pm.004$ & $.711\pm.018$ \\
      & Weighted $\ell_1$ & 8 & $.760\pm.003$ & $.758\pm.018$ \\
      & Conditional BT & 224 & $.802\pm.003$ & $\mathbf{.777\pm.020}$ \\
    \midrule
    Climate, both cues
      & Attribute match count & 0 & $.635\pm.002$ & $.635\pm.012$ \\
      & Party cue only & 0 & $.596\pm.002$ & $.594\pm.014$ \\
      & AMCE-ordered tree & 9 & $.658\pm.003$ & $.656\pm.017$ \\
      & Weighted $\ell_1$ & 9 & $.692\pm.003$ & $.687\pm.015$ \\
      & Conditional BT & 280 & $.749\pm.004$ & $\mathbf{.710\pm.018}$ \\
    \bottomrule
  \end{tabular}
\end{table*} 

\subsection{Participant-Held-Out Evaluation}

Table~\ref{tab:lp-results} reports results for all three task environments.
The General and concern-only Climate tasks each contain 5,755 comparisons from
1,151 respondents, while the both-cues Climate task contains 5,760 comparisons
from 576 respondents.
Because the task remits, displayed cue sets, and Survey~2 sample differ,
absolute accuracy differences across task environments are descriptive.

We construct 20 reproducible 85/15 train--test splits by respondent, not by
comparison. Exact prediction ties receive half
credit, corresponding to random choice. Because every respondent
contributes five direct comparisons, comparison-level accuracy and
respondent-averaged accuracy coincide. Table~\ref{tab:lp-results} reports the
resulting training and test accuracies.
In the concern-only Climate task, the learned weighted $\ell_1$
metric correctly predicts at least four of five choices for 64.7\% of held-out
respondents. The
supplementary
material reports the full participant-level distributions for all three
tasks.

\subsection{What the Comparisons Reveal}

The cue-only model trails weighted $\ell_1$ in all three
tasks, showing that the largest AMCE does not summarize whole-profile choice.
Other displayed attributes retain predictive value even when one political or
topic-specific cue is especially salient; using only the largest training-set
AMCE therefore leaves predictive information unused.

Weighted $\ell_1$ outperforms both equal match count and the AMCE-ordered tree in every task. Thus, equal weighting is not an adequate predictive summary,
consistent with the unequal mismatch penalties in the AMCE analysis. The
weaker performance of the AMCE-ordered tree likewise shows that a single,
shared lexicographic rule~\cite{gigerenzer1996reasoning} is not sufficient at
the aggregate level. Across tasks, choices are better summarized by a compact
model that permits unequal weights, ordinal distances, and compensatory
trade-offs among attributes.

Conditional BT attains the highest held-out accuracy in
every task. Its additional respondent-conditioned flexibility is therefore
predictively useful, although this comparison does not isolate which added
terms produce the gain. This result is consistent with the descriptive
subgroup patterns in Figure~\ref{fig:app-filtered-age}, including larger
age-mismatch penalties among the youngest respondents and larger
gender-mismatch penalties among female respondents. Its advantage
is modest relative to its greater complexity, however, and its larger train--test gap is consistent with some overfitting.
Weighted $\ell_1$ therefore provides a parsimonious, interpretable
approximation, while conditional BT captures a smaller residual layer of
profile-conditioned heterogeneity.
The participant-level distributions tell a similar story in the concern-only Climate task: weighted $\ell_1$ predicts at least four of five choices correctly for 64.7\% of held-out respondents, compared with 67.5\% for conditional BT. The supplementary material reports the full distributions for all three tasks.

\paragraph{Party and concern provide complementary predictive
information.} 
Survey~2 presents political alignment and climate concern together, allowing us to ask whether their predictive contributions overlap. Holding the seven demographic attributes fixed, weighted $\ell_1$ achieves held-out accuracy of .655 with party, .646 with concern, and .687 with both. Conditional BT shows the same pattern, increasing from .670 with party and .658 with concern to .710 with both. For both model families, the higher
accuracy when both cues are included is consistent with each supplying
predictive information not contained in the other. The supplementary material
reports the demographics-only specifications and test log loss.

    \section{Limitations and Future Work}

\subsection{Limitations}

Several limitations bound these conclusions. The conjoint tasks elicit choices between synthetic individual profiles rather than evaluations of complete panel compositions. An alternative design could ask respondents to compare panels based on summary charts of their composition. However, such comparisons require integrating multiple quantitative trade-offs across several dimensions simultaneously, making them a more demanding  cognitive task. Our design instead isolates this individual-level judgment of perceived representation. How such judgments aggregate into evaluations of complete panels or their legitimacy remains an important direction for future work. More broadly, these design choices bound the scope of inference. The randomized design identifies effects of the displayed attributes within the experimental profile distribution; it does not establish how those effects generalize to omitted attributes or profile distributions encountered in practice. Cross-task comparisons are therefore descriptive, subgroup and value-specific patterns remain exploratory, and the two UK samples limit broader generalizability. The quota experiment has a similarly bounded interpretation: its qualitative conclusion is stable across the panel sizes, quota tolerances, and selection procedures we test, but the exact magnitudes remain specific to the observed pool and tested configurations.

Predictive accuracy measures generalization within these survey environments,
not democratic legitimacy or welfare; residual errors leave substantial room
for unobserved considerations and individual variation. More broadly,
perceived representation is only one assembly objective. Descriptive
representativeness, substantive inclusion, deliberative quality, and public
legitimacy can diverge~\citep{oecd2020deliberative,pow2020likeme}. Optimizing
similarity alone could reproduce homophily, so any operational use should
retain explicit inclusion safeguards and treat predictive scores as diagnostic
rather than as a stand-alone selection objective.

\subsection{Future Work}
Future work should replicate these experiments across countries, policy domains, and citizens' assembly remits, and test the broader range of attributes identified through the open-ended responses. The observed heterogeneity across respondent groups also motivates studying why different populations place different weights on particular characteristics, and whether these differences become more pronounced across cultural and institutional contexts. Future work should also investigate how individual judgments of representation aggregate into evaluations of complete panels, both before and after deliberation, and whether these judgments change as participants gain experience with the deliberative process. Finally, future work should develop panel-selection procedures that optimize empirically estimated models of perceived representation while simultaneously satisfying explicit diversity, inclusion, and fairness constraints.

\bibliography{main}

\onecolumn
\appendix
\section{Further Related Work}
\label{sec:appendix-related-work}

\paragraph{Citizens' assemblies, sortition, and deliberative democracy.}
The theory of deliberative democracy argues that democratic legitimacy requires
informed public deliberation rather than mere aggregation of
preferences~\citep{fishkin2018democracy, fishkin2005experimenting}. The global
``deliberative wave'' has seen the proliferation of citizens' assemblies across
diverse contexts~\citep{oecd2020deliberative}, and public support for these
sortition-based bodies has been documented across multiple
countries~\citep{pilet2023public}. A key algorithmic challenge is balancing equal
selection probability with demographic representativeness:
\citet{flanigan2021fair} formalized this tension and introduced the LEXIMIN
algorithm, while \citet{flanigan2024transparent} extended this work to
incorporate transparency and manipulation-robustness.
\citet{ebadian2022sortition} introduced a metric-based representation framework as a tool
for analyzing whether panels  represent not only individual attributes but intersectional as well. More
recently, permanent citizens' assemblies, such as the Ostbelgien Model in
Belgium and experimental formats in the European Union, have further heightened
the importance of principled panel selection, as panels are periodically rotated
through sortition.
Recent algorithmic work has addressed practical deployment challenges:
\citet{assos2025alternates} study optimal alternate selection to maintain
representativeness after panelist dropout, and
\citet{gambhir2025dropout} provide near-optimal dropout-robust sortition
algorithms with formal guarantees.
Beyond assembly design, \citet{sana2026dempo} connect sortition to AI
alignment by using sortition-weighted preference data to train language models
that reflect demographically representative preferences.

\paragraph{Metric-based proportional representation and fair selection.}
Metric distortion research studies how well a small committee can represent a
population when preferences are drawn from a 
metric space where all the individuals are co-located~\citep{anshelevich2018approximating, caragiannis2024metric}, and how
pairwise comparison queries can reduce
distortion~\citep{ebadian2024pairwise}. \citet{aziz2024proportional} introduced
proportional fairness in metric spaces, showing that the Expanding Approvals
Rule achieves constant-factor proportional fairness using only ordinal
information. The FairGreedyCapture algorithm of \citet{ebadian2025boosting}
achieves both fairness and constant-factor core approximation, connecting to the
broader literature on proportionally fair
clustering~\citep{chen2019proportionally, micha2020proportionally}. Our work
complements these theoretical contributions by providing empirical preference
data that can inform the construction and validation of the metric spaces these
algorithms assume.

\paragraph{Metric learning from preferences.}
Metric learning aims to learn a distance function from data, typically in the
form of a Mahalanobis
metric~\citep{xing2002distance, weinberger2009distance}. \citet{kulis2013metric}
provides a comprehensive survey of the field, and \citet{xu2024metric} studies
metric learning from limited pairwise preference comparisons.
Our
evaluation uses direct pairwise choices with repeated participant-level
train--test splits, so no respondent appears in both partitions of a split.

\section{Ethics Approval}
\label{sec:appendix-ethics}

This study was reviewed by the University of Southern California Institutional Review Board and determined to be exempt under §46.104(d)(2) (Study ID: UP-25-00989).

\section{Disclosure of AI Usage}
\label{sec:appendix-ai-disclosure}

This paper was written with the assistance of Claude (Sonnet 4.5 and Opus 4.6, Anthropic) accessed through Claude Code, an agentic command-line coding tool. We disclose the following uses. All authors take full responsibility for the correctness and veracity of all content in this paper.

\paragraph{Literature discovery.}
The AI was used to search for and summarize potentially relevant prior work that may have been outside the authors' immediate awareness. All identified papers were independently evaluated by the authors for relevance and accuracy before inclusion; no citation was added without author verification.

\paragraph{Prose rewriting and polishing.}
Author-written drafts were provided to the AI for rewriting to improve clarity, readability, and terminological consistency. In all cases, the initial draft was written by the authors; the AI was used to refine wording and correct grammar. The authors reviewed all resulting text for correctness and faithfulness to the intended content.

\paragraph{Analysis code.}
The AI assisted with writing and debugging data analysis scripts (e.g., AMCE estimation, metric learning pipelines, and plotting). All analysis logic and experimental design decisions were made by the authors; the AI served as a coding assistant. The authors reviewed and verified all AI-produced code to the best of their knowledge and ability.

\paragraph{Scope limitations.}
The AI was \emph{not} used to generate any core ideas, research questions, novel arguments, experimental designs, formal definitions, mathematical content, tables, or figures. The AI did not suggest structural changes to the paper (section ordering, argument flow, or paper organization). All intellectual contributions are entirely the work of the authors.

\section{Experiment Details}
\label{sec:appendix-data-collection}
\label{sec:appendix-experiment-details}

\subsection{Fielding, Screening, and Analysis Samples}

Survey~1 was fielded on Prolific in four UK waves,
each targeting approximately 300 completed respondents. It contained the
general-purpose task with political alignment and the climate-focused task with
climate concern; the same respondents completed both tasks. Of 1,228 raw
records, we excluded 14 non-UK records and 15 records without a completion
time, leaving 1,199 otherwise-valid responses. Their 2nd and 98th completion-time
percentiles are 3.080 and 17.751 minutes. We excluded 24 responses below the
lower cutoff and 24 above the upper cutoff, leaving 1,151 respondents.

Survey~2 was fielded in two UK waves. Of 609 raw records, we excluded 7 incomplete survey responses, leaving 602
otherwise-valid responses. Completion time is computed directly from the
stored start and completion timestamps. The 2nd and 98th percentiles are
1.942 and 10.000 minutes. We excluded 13 responses below the lower cutoff and
13 above the upper cutoff, leaving 576 respondents: 288 from each wave. In both
surveys, percentiles are computed over the
otherwise-valid records using linear interpolation, and observations equal to
a cutoff are retained.

\begin{table}[hbt!]
  \centering
  \small
  \caption{Survey tasks and analysis counts. A five-candidate ranking supplies
  all $\binom{5}{2}=10$ implied winner--loser relations. Ranking-derived
  relations are not used in the AMCE analysis.}
  \label{tab:appendix-survey-roster}
  \begin{tabular}{lrrr}
    \toprule
    Task environment & $N$ & Direct choices & Rank pairs \\
    \midrule
    General-purpose & 1,151 & 5,755 & 23,020 \\
    Climate-focused, concern only & 1,151 & 5,755 & 23,020 \\
    Climate-focused, both cues & 576 & 5,760 & 0 \\
    \bottomrule
  \end{tabular}
\end{table}

The
surveys shared two core components: respondents reported the attributes used
to construct their profile and then chose the more representative member from
randomized pairs. Survey~1 presented five pairs per task environment and also
included two five-candidate rankings per environment. In each ranking,
respondents ordered profiles from 1 (``represents you most'') to 5
(``represents you least''), supplying the ten implied winner--loser relations.
Survey~1 then invited general feedback on the survey. Survey~2 instead
presented ten pairs and ended with the targeted question about additional
attributes used in the open-ended elicitation analysis.
Table~\ref{tab:appendix-survey-roster} reports the resulting quantitative
analysis counts.

\paragraph{Sample composition.}
Table~\ref{tab:appendix-sample-composition} reports the complete composition
of both analysis samples on all nine recorded attributes. The two surveys are
separate samples, so comparisons across their task environments are descriptive
rather than a randomized effect of changing the displayed cue set.

\begin{table}[h!]
  \centering
  \small
  \caption{Composition of the two analysis samples. Counts sum to the
  corresponding analysis-sample size within each attribute. Survey~1
  respondents appear in both Survey~1 task environments. Solid rules separate
  attributes.}
  \label{tab:appendix-sample-composition}
  \begin{tabular}{p{0.12\textwidth}p{0.39\textwidth}p{0.39\textwidth}}
    \toprule
    Attribute & Survey 1 ($N=1{,}151$) & Survey 2 ($N=576$) \\
    \midrule
    \midrule
    Gender &
    Female 603; Male 542; Other 6 &
    Female 301; Male 273; Other 2 \\
    \midrule
    Age group &
    16--29: 200; 30--44: 304; 45--60: 375; 60+: 272 &
    16--29: 101; 30--44: 151; 45--60: 196; 60+: 128 \\
    \midrule
    Area type &
    Rural 212; Suburban 653; Urban 286 &
    Rural 104; Suburban 314; Urban 158 \\
    \midrule
    Ethnicity &
    Asian 68; Black 56; Mixed 28; Other 11; White 988 &
    Asian 43; Black 45; Mixed 13; Other 1; White 474 \\
    \midrule
    Disability &
    No 850; Prefer not to say 25; Yes 276 &
    No 410; Prefer not to say 8; Yes 158 \\
    \midrule
    Education &
    A-level 252; Apprenticeship 60; Bachelor's 473; GCSE 133;
    No qualifications 8; Postgraduate 225 &
    A-level 125; Apprenticeship 30; Bachelor's 230; GCSE 66;
    No qualifications 4; Postgraduate 121 \\
    \midrule
    Region &
    East Midlands 102; East of England 84; London 147; North East 56;
    North West 137; Northern Ireland 22; Scotland 89; South East 149;
    South West 109; Wales 49; West Midlands 102; Yorkshire and the Humber 105 &
    East Midlands 43; East of England 45; London 64; North East 26;
    North West 72; Northern Ireland 9; Scotland 43; South East 80;
    South West 55; Wales 27; West Midlands 51; Yorkshire and the Humber 61 \\
    \midrule
    Party &
    Conservative 176; Green 156; Labour 355; Liberal Democrat 121;
    Other/None 132; Plaid Cymru 10; Reform UK 172; SNP 29 &
    Conservative 105; Green 55; Labour 229; Liberal Democrat 54;
    Other/None 25; Plaid Cymru 4; Reform UK 85; SNP 19 \\
    \midrule
    Climate concern &
    Fairly 501; Not very 172; Not at all 43; Very 435 &
    Fairly 272; Not very 60; Not at all 17; Very 227 \\
    \bottomrule
  \end{tabular}
\end{table}
\FloatBarrier

\subsection{Participant-Facing Citizens' Assembly Context}
\label{sec:appendix-participant-context}

\begin{figure}[hbt!]
  \centering
  \includegraphics[width=0.72\textwidth]{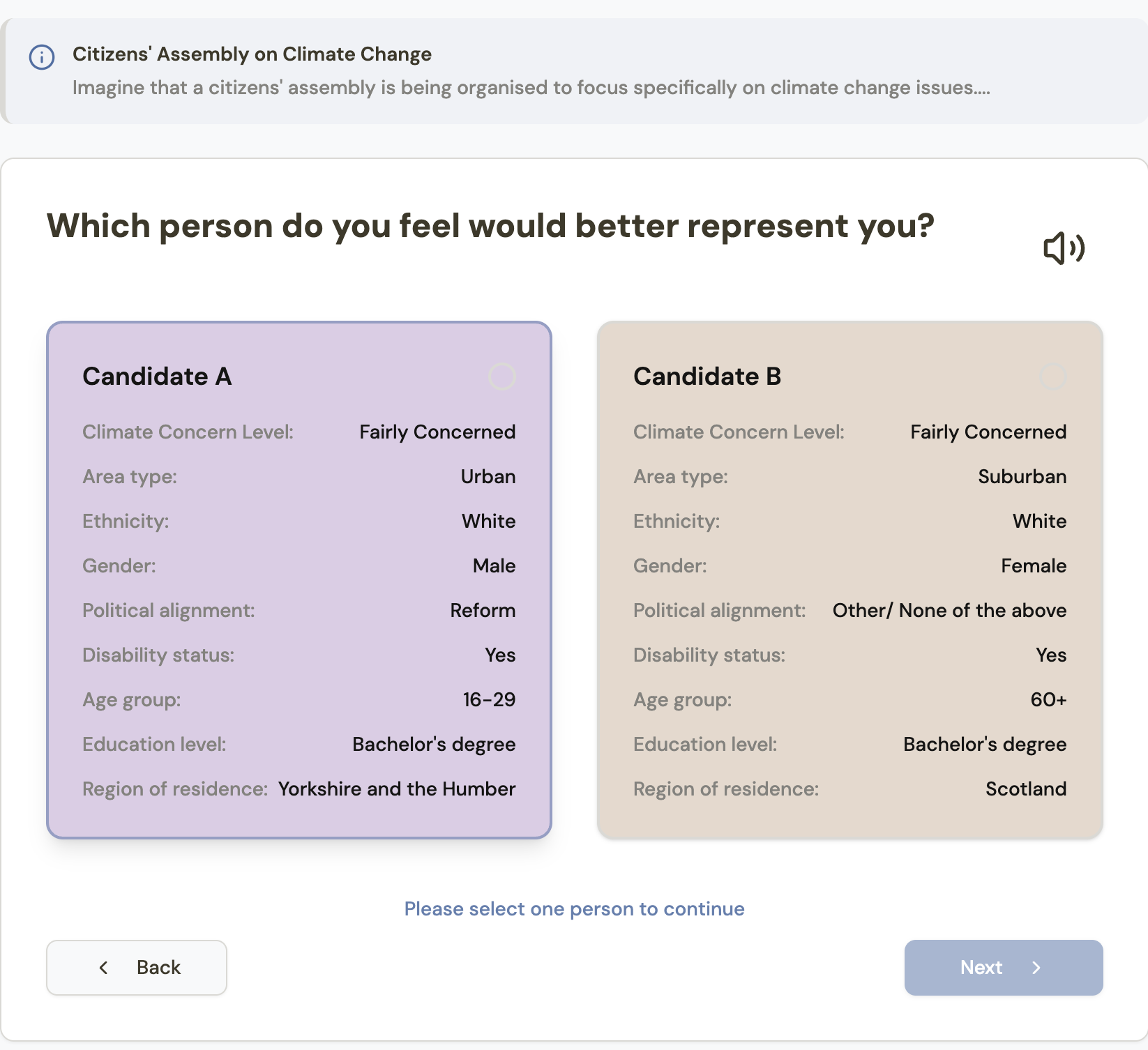}
  \caption{Participant-facing pairwise-choice interface for Survey~2's
  climate task, displaying both political alignment and climate concern.}
  \Description{The Survey 2 both-cues climate pairwise-choice interface,
  displaying two nine-attribute candidate profiles.}
  \label{fig:appendix-pairwise-interface}
\end{figure}

Both surveys introduced citizens' assemblies and the judgment task before
eliciting choices. Their landing pages made available the same short
research-team-produced explainer video and gave the following written
introduction:
\begin{quote}
\small
Citizens' assemblies are a way for everyday people to take part directly in
important public decisions. In a citizens' assembly, a group of people is
chosen at random from the community. They are not politicians or experts. They
are people like us---shopkeepers, teachers, carers, tradespeople, doctors,
students, friends, and family members---who come together to tackle a shared
problem. The idea is simple: if decisions affect everyone, then every
perspective should be represented. This means that even people who do not take
part directly can still be represented by someone with a similar background
and lived experiences.
\end{quote}
The welcome screen then stated the purpose and sequence of the survey. Survey~1
explained that respondents would consider two citizens' assemblies, one
addressing various issues and one focused on climate change. Survey~2 described
one climate-focused assembly and previewed the ten pairwise choices. 
Figure~\ref{fig:appendix-pairwise-interface}
shows the Survey~2 pairwise-comparison interface.

After answering the background questions, respondents encountered a dedicated
instruction screen immediately before each task environment and had to select
\emph{Continue} to advance. Table~\ref{tab:appendix-context-wording} reproduces
the context displayed on these screens.

\begin{table}[hbt!]
  \centering
  \small
  \caption{Written context displayed immediately before each set of
  representation choices.}
  \label{tab:appendix-context-wording}
  \begin{tabular}{p{0.20\textwidth}p{0.70\textwidth}}
    \toprule
    Survey and task & Participant-facing instruction \\
    \midrule
    Survey~1, general-purpose &
    Imagine that a citizens' assembly is being organised to recommend
    solutions to a variety of issues. In the following questions, you will be
    asked to choose which person you think would better represent you, as
    someone who shares a similar background, experiences and concerns. \\
    \addlinespace
    Survey~1, climate-focused &
    Now, imagine that a citizens' assembly is being organised to focus
    specifically on climate change issues. Again, you will be asked to choose
    which person you think would better represent you, as someone who shares a
    similar background, experiences, and concerns. Please keep in mind that
    this assembly is focused on climate change. \\
    \addlinespace
    Survey~2, climate-focused &
    Imagine that a citizens' assembly is being organised to focus specifically
    on climate change issues. You will be asked to choose which person you
    think would better represent you, as someone who shares a similar
    background, experiences, and concerns. \\
    \bottomrule
  \end{tabular}
\end{table}

Finally, every pairwise screen repeated the criterion: ``Which person do you
feel would better represent you, as someone who shares a similar background,
experiences, and concerns?'' Thus, the broad definition, the assembly-specific
remit, and the intended meaning of the comparison were supplied at distinct
points in the survey.

The video was click-to-play: playback was neither required nor recorded, and
the surveys did not include a citizens' assembly comprehension check. We can
therefore document the information made available and repeated in writing, but
not each respondent's prior familiarity or whether every respondent watched
the video.

\subsection{Candidate-Profile Randomization}

For the seven shared attributes, and for the single political or climate cue
shown in each Survey~1 task, a candidate matches the respondent independently
with probability $1/2$ whenever the respondent's value is allowed in candidate
profiles. Conditional on mismatch, the candidate value is drawn uniformly from
the remaining candidate levels. Candidate profiles exclude ``Prefer not to
say'' for disability and ``Other'' for ethnicity; respondents who report these
values therefore always receive a mismatch on the corresponding attribute and
do not contribute to that attribute's mismatch AMCE. The two candidate
positions are exchangeable because the candidates are generated
independently by the same procedure. The displayed attribute order is
randomized once and held fixed within respondent. For all candidate-admissible
self-values, this construction equalizes the match probability across
respondents whose own category frequencies differ.

Survey~2 displays political alignment and climate concern together. To avoid
generating these two values independently, its sampler uses the empirical
party-by-concern count matrix from the pre-screening UK Survey~1 response pool.
This matrix contains all 1,214 UK records with both self-reports and was fixed
before Survey~2 was fielded. It is therefore distinct from the 1,151-person
cleaned analysis pool used for the AMCE analysis. 

\begin{table}[hbt!]
  \centering
  \small
  \caption{Pre-screening UK Survey~1 party-by-climate-concern counts used by
  the correlated Survey~2 profile sampler ($N=1{,}214$).}
  \label{tab:appendix-sampler-party-concern-matrix}
  \begin{tabular}{lrrrr}
    \toprule
    & Fairly & Not very & Not at all & Very \\
    \midrule
    Conservative & 87 & 42 & 4 & 50 \\
    Green & 51 & 2 & 1 & 113 \\
    Labour & 168 & 35 & 4 & 165 \\
    Liberal Democrat & 71 & 11 & 0 & 45 \\
    Other/None & 64 & 32 & 9 & 36 \\
    Plaid Cymru & 3 & 2 & 0 & 6 \\
    Reform UK & 71 & 56 & 26 & 27 \\
    SNP & 18 & 2 & 3 & 10 \\
    \bottomrule
  \end{tabular}
\end{table}
\FloatBarrier

Writing this matrix as $C$, the sampling distribution is
\[
 \widetilde p_{rs}
 =0.95\frac{C_{rs}}{1{,}214}+0.05\frac{1}{8\cdot4}.
\]
For each candidate, the sampler first draws party-match and concern-match
indicators independently from $\operatorname{Bernoulli}(1/2)$. It then removes
all cells of $\widetilde p$ inconsistent with those two decisions, renormalizes
the remaining cells, and draws the candidate's party--concern pair. Each of the
four match--mismatch states therefore has probability $1/4$, while values
within a state retain the smoothed empirical association. Smoothing gives every
party--concern cell positive support. For example, for a respondent who
supports Labour and is Very Concerned, the state ``party mismatch, concern
match'' restricts the draw to non-Labour cells in the Very Concerned column;
the probabilities of those cells are then renormalized before one is drawn.
Thus, the two match indicators determine which of four blocks is used, while
the smoothed empirical matrix determines the particular values drawn within
that block. This procedure is applied independently
to the two candidates in each comparison. 

\subsection{Cleaning and Reproducibility}

The cleaning pipeline applies the same 2nd--98th percentile duration rule
described above and then maps both surveys' heterogeneous raw field names to the
canonical attributes used throughout the analysis. It discards browser and
Prolific identifiers, assigns stable pseudonyms only to preserve within-survey
linkage, and validates task names, candidate identifiers, and
condition-specific attributes. Missing values remain missing rather than being
coded as matches. Direct choices and ranking-derived comparisons carry
separate task-source labels. Source hashes, exclusion counts, condition
schemas, random seeds, and machine-readable analysis tables are stored with the
analysis pipeline.

\paragraph{Ethics and consent.}
Participants provided informed consent through Prolific.

\section{AMCE Analysis Details}
\label{sec:amce-appendix}

\subsection{Estimator and Uncertainty}

\paragraph{Candidate-row outcomes.}
The AMCE analysis uses only direct pairwise choices.
We represent each
choice as two candidate rows, one for each displayed profile.
Let $j$ index these candidate rows for respondent $i$, and define
$Y_{ij}=1$ if that candidate was selected and $Y_{ij}=0$ otherwise. Thus, the
two rows from any one choice have outcomes 1 and 0. For displayed attribute
$a$, let $p_{ia}$ be respondent $i$'s value and
$c_{ja}$ the value shown in candidate row $j$, and define
$M_{ija}=1\{c_{ja}\ne p_{ia}\}$. Thus, $M_{ija}=0$ denotes a match
and $M_{ija}=1$ a mismatch.

\paragraph{Respondent-level contrast.}
For each state $m\in\{0,1\}$, collect respondent $i$'s candidate rows
with that state in
$\mathcal J_{ia}(m)=\{j:M_{ija}=m\}$. The respondent's mean selection outcome
in that state is
\[
 \overline Y_{ia}(m)
 =\frac{1}{|\mathcal J_{ia}(m)|}
   \sum_{j\in\mathcal J_{ia}(m)}Y_{ij}.
\]
Consequently, $\overline Y_{ia}(1)$ is the fraction of respondent
$i$'s mismatching candidate rows that were selected, whereas
$\overline Y_{ia}(0)$ is the corresponding fraction among matching rows.
We define respondent $i$'s
mismatch effect for attribute $a$ as
\[
 \widehat\tau_{ia}
 =\overline Y_{ia}(1)-\overline Y_{ia}(0).
\]
A negative value means that respondent $i$ selected mismatching
candidates less often than matching candidates; for example, an effect of
$-0.10$ is a ten-percentage-point difference in selection rates.

\paragraph{Across-respondent estimate.}
A respondent contributes an
effect for attribute $a$ only when both $\mathcal J_{ia}(0)$ and
$\mathcal J_{ia}(1)$ are nonempty. Let $\mathcal I_a$ be the set of these
eligible respondents, and let $N_a=|\mathcal I_a|$. The reported
mismatch AMCE is the arithmetic mean of their respondent-level effects:
\[
 \widehat\tau_a
 =\frac{1}{N_a}\sum_{i\in\mathcal I_a}\widehat\tau_{ia}.
\]
A negative
$\widehat\tau_a$ therefore means that, for the average eligible respondent, a
mismatching candidate is selected less often than a matching candidate.
Because the candidate rows are
averaged within respondent before respondents are averaged, every eligible
respondent receives equal weight regardless of how many usable rows they
contribute.

\paragraph{Uncertainty.}
Let $s_a$
be the sample standard deviation of
$\{\widehat\tau_{ia}:i\in\mathcal I_a\}$. The standard error of the mean is
$s_a/\sqrt{N_a}$. The reported 95\% interval is
\[
 \widehat\tau_a
 \ \pm\ t_{0.975,N_a-1}\,
 \frac{s_a}{\sqrt{N_a}}.
\]
Here $t_{0.975,N_a-1}$ is the 97.5th percentile of a Student
$t$ distribution with $N_a-1$ degrees of freedom. The eligible respondent
count can differ slightly by attribute because a respondent must have observed
both match and mismatch states. The focal
match--mismatch state is randomized independently of the other attribute
mismatch states.

\subsection{Exploratory Subgroup and Value-Specific Analyses}

We use two complementary exploratory analyses to examine variation
hidden by the overall mismatch AMCEs. Filtered AMCEs ask whether the
mismatch effects differ across respondent subgroups.
Value-specific
heatmaps ask how respondents with a given attribute value react to particular
alternative values, rather than combining every mismatch into one contrast.

\paragraph{Filtered AMCE estimator.}
\label{sec:appendix-segments-additional}

For a filtered analysis, we first restrict the data to
one respondent subgroup and then repeat the respondent-first AMCE estimator
defined above. For example, the age-filtered analysis separately
estimates every displayed attribute's mismatch effect among respondents aged
16--29, 30--44, 45--60, and 60+. We require at least 30 respondents in a
subgroup and at least 30 eligible respondents for an estimate. These subgroup
analyses are descriptive. The confidence intervals apply to individual
estimates, and we do not formally test whether effects differ across groups or
adjust for the number of comparisons examined. We therefore use the results to
identify possible patterns, not to make definitive claims about subgroup
differences.

\paragraph{Age- and gender-filtered results.}
Age and gender show the clearest
recurring descriptive patterns. 
Figure~\ref{fig:app-filtered-age} in the main text
shows these two filtered analyses across all three task environments;
Table~\ref{tab:appendix-age-gender-amce}
below reports the corresponding focal estimates numerically.

\begin{table}[H]
  \centering
  \small
  \setlength{\tabcolsep}{4pt}
  \caption{Exploratory subgroup AMCEs for gender and age similarity. Rows
  divide respondents by self-reported gender or age. Each cell reports the
  candidate-mismatch effect for that same attribute in the indicated task
  environment, with its pointwise 95\% confidence interval and eligible 
  subgroup size $N$ (respondents who observed both match and mismatch
  states). Negative estimates indicate a preference for candidates who share the
  respondent's attribute value.}
  \label{tab:appendix-age-gender-amce}
  \begin{tabular}{l|l|r|r|r}
    \toprule
    Filter & Group & General-purpose & Climate, concern only & Climate, both cues \\
    \midrule
    Gender & Female &
    $-.124\;[-.152,-.095]$ ($N=597$) &
    $-.098\;[-.127,-.069]$ ($N=600$) &
    $-.063\;[-.092,-.033]$ ($N=301$) \\
    Gender & Male &
    $-.046\;[-.078,-.013]$ ($N=542$) &
    $-.031\;[-.060,-.001]$ ($N=542$) &
    $-.001\;[-.029,.027]$ ($N=273$) \\
    Age & 16--29 &
    $-.208\;[-.253,-.162]$ ($N=200$) &
    $-.168\;[-.221,-.115]$ ($N=200$) &
    $-.121\;[-.172,-.071]$ ($N=101$) \\
    Age & 30--44 &
    $-.118\;[-.158,-.077]$ ($N=303$) &
    $-.100\;[-.140,-.061]$ ($N=304$) &
    $-.069\;[-.108,-.029]$ ($N=151$) \\
    Age & 45--60 &
    $-.119\;[-.156,-.081]$ ($N=375$) &
    $-.098\;[-.137,-.060]$ ($N=374$) &
    $-.063\;[-.099,-.027]$ ($N=196$) \\
    Age & 60+ &
    $-.103\;[-.148,-.058]$ ($N=271$) &
    $-.100\;[-.140,-.059]$ ($N=271$) &
    $-.086\;[-.130,-.042]$ ($N=128$) \\
    \bottomrule
  \end{tabular}
\end{table}
\FloatBarrier

\FloatBarrier

The main text discusses the recurring age- and gender-filtered patterns. As
noted above, these are descriptive subgroup comparisons rather than formal
interaction tests. The remaining filtered analyses did not reveal additional recurring
descriptive patterns sufficiently clear to warrant focused discussion. We
omit those plots, although they can be regenerated from the released data and
source code. Their omission should not be interpreted as evidence that no
other subgroup heterogeneity exists.

\paragraph{Value-specific heatmaps.}
\label{sec:appendix-interaction-heatmaps}

\begin{figure}[!htbp]
  \centering
  \includegraphics[width=0.564\textwidth,height=0.48\textheight,keepaspectratio]{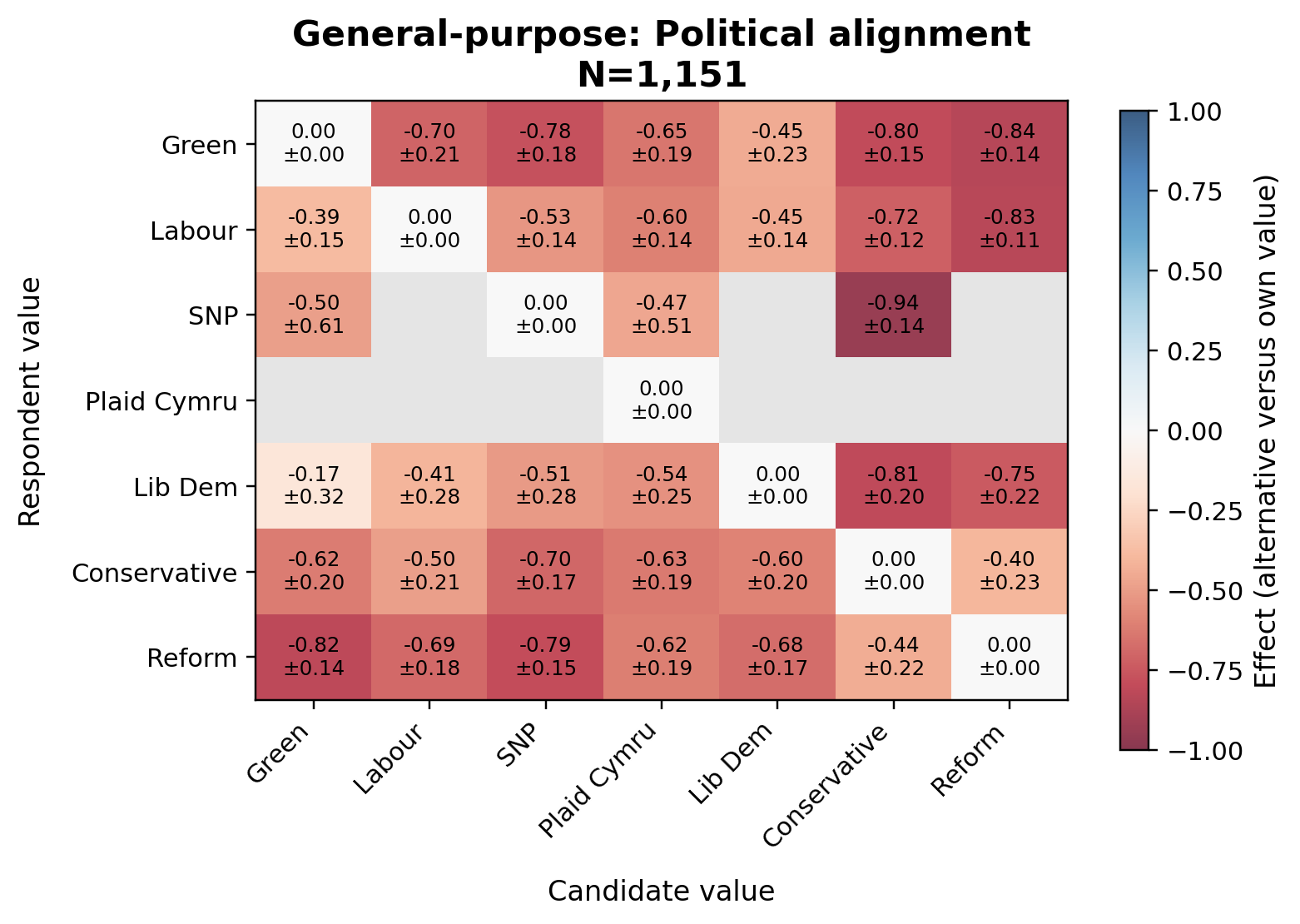}
  \caption{Group-conditioned own-versus-alternative effects for political
  alignment in the general-purpose task. Each cell reports
  $2\Pr(\text{column-party candidate chosen})-1$, averaged first within
  respondent; text gives the estimate and pointwise 95\% interval half-width.
  Grey cells have fewer than ten supporting respondents.}
  \Description{Heatmap of respondent-averaged own-versus-alternative choice
  effects across political parties in the general-purpose task.}
  \label{fig:app-heatmap-general-politics}
\end{figure}

Value-specific analyses compare, for respondents with value $\ell$, a candidate
who matches $\ell$ with a candidate carrying a particular alternative
$\ell'$. For each ordered pair $(\ell,\ell')$, we retain only
direct choices in which one candidate has the respondent's own value $\ell$
and the other has the alternative $\ell'$. Let
$\mathcal I_{\ell,\ell'}$ contain respondents with own value $\ell$ who
encounter at least one such choice, and let
$N_{\ell,\ell'}=|\mathcal I_{\ell,\ell'}|$. If
$\overline Y_i(\ell')$ is respondent $i$'s mean indicator for choosing the
$\ell'$ candidate in those choices, the plotted effect is
\[
  \frac{1}{N_{\ell,\ell'}}
  \sum_{i\in\mathcal I_{\ell,\ell'}}
  \left(2\overline Y_i(\ell')-1\right).
\]
Thus, zero denotes an even split, negative values favor the respondent's own
value, and positive values favor the displayed alternative. We average within
respondent before averaging across respondents and construct pointwise
$t$-intervals from these respondent-level effects. Diagonal cells are fixed at
zero by definition; cells supported by fewer than ten respondents are omitted.
The main text reports the selected age
and climate-concern heatmaps. Figure~\ref{fig:app-heatmap-general-politics}
below reports political alignment in the Survey~1 general-purpose task. Each
heatmap displays respondent values in rows and alternative candidate values in
columns; cells with fewer than ten supporting respondents are greyed out.

\FloatBarrier

Like the age and climate-concern heatmaps in
Figure~\ref{fig:app-heatmap-general-a}, the political-alignment heatmap in
Figure~\ref{fig:app-heatmap-general-politics} shows that a binary
match--mismatch effect conceals variation across particular respondent and
candidate values.

For the nominal party attribute, the point estimates vary
by party pairing and direction rather than following an ordinal gradient:
Labour and Conservative respondents respond differently to Green and Reform
alternatives, and the estimate for one party's supporters evaluating another
need not match the reverse comparison. These descriptive patterns suggest that
political mismatches are not interchangeable. They also illustrate a
limitation of the weighted-$\ell_1$ model: one political-alignment weight
cannot distinguish every respondent--candidate party pairing. Grey cells
contain fewer than ten supporting respondents. The remaining value-specific
heatmaps did not reveal additional descriptive patterns sufficiently clear to
warrant focused discussion. They can be regenerated from the released data
and source code; their omission is not evidence that no other value-specific
heterogeneity exists.

\section{Open-Ended Attribute Elicitation and Survey Feedback}
\label{sec:appendix-open-ended}

\paragraph{Survey~1 feedback.}
Both surveys ended with an optional free-text item, but the questions served
different purposes. Survey~1 invited general comments through the prompt quoted
in the main text. The two responses discussed there raised concerns about
demographic stereotyping and about the inability of static profile attributes
to capture how a representative reasons and decides. Because the question
solicited general feedback rather than additional attributes, we treat these
comments as illustrative and do not include them in the Survey~2 coding or
counts below.

\paragraph{Survey~2 question and sample.}
Survey~2 instead asked respondents to suggest attributes that might help
measure representation but were absent from the displayed profiles. The main
text gives the wording used in the first wave. In the second wave, the prompt
inserted ``not covered above'' after ``other attributes'' but was otherwise
unchanged. Each wave contributed 288 respondents to the cleaned Survey~2
sample. Across the two waves, 261 respondents supplied a non-empty answer and
315 left the item blank.

\paragraph{Survey~2 coding.}
We assigned one or more descriptive themes to each non-empty response; 82 of
the 261 responses received multiple labels. The coding began from existing
manual and deterministic keyword assignments. We then reviewed every response
and corrected assignments for which negation or context made a keyword match
misleading. Responses that only restated political alignment or climate
concern, both of which were already displayed in Survey~2, were retained in the
response-level record but excluded from the inventory of additional themes.
Likewise, ``no additional attribute'' was coded separately. Seven respondents
used that label while also emphasizing at least one attribute, so these
categories overlap.

Browser identifiers and timestamps were removed from the coding record. The
counts in Table~\ref{tab:appendix-open-ended} are respondent-level mentions;
they are neither mutually exclusive categories nor model-based estimates. The
deidentified responses, reviewed theme assignments, and aggregate counts are
included in the supplementary materials.

\begin{table}[H]
  \centering
  \scriptsize
  \setlength{\tabcolsep}{4pt}
  \caption{Themes in Survey~2 open-ended responses after omitting repetitions
  of political alignment and climate concern. Counts are respondent-level
  mentions under multi-label coding; percentages use the 261 non-empty answers
  as the denominator and therefore do not sum to 100\%.}
  \label{tab:appendix-open-ended}
  \begin{tabular}{@{}p{0.22\textwidth}rrp{0.52\textwidth}@{}}
    \toprule
    Theme & Count & \% & Included responses \\
    \midrule
    Deliberative qualities & 27 & 10.3 & Listening, openness, honesty, respect, empathy, fairness, or willingness to deliberate. \\
    Socioeconomic position & 22 & 8.4 & Income, wealth, class, financial circumstances, benefits, or housing tenure. \\
    Occupation or work experience & 20 & 7.7 & Employment status, sector, industry, occupation, or work experience. \\
    Family or household status & 19 & 7.3 & Children, parenthood, marital status, or household composition. \\
    Local or community connection & 19 & 7.3 & Local residence, community roots, service, accessibility, or community involvement. \\
    General capability or education & 19 & 7.3 & Education, intelligence, communication ability, competence, or general experience. \\
    Other policy views & 18 & 6.9 & Substantive views outside climate, including immigration and welfare. \\
    Past environmental action or involvement & 18 & 6.9 & Personal climate action, recycling, petitions, activism, initiatives, or group involvement. \\
    Climate knowledge or evidence orientation & 17 & 6.5 & Scientific or environmental knowledge, relevant expertise, or evidence-based reasoning. \\
    Personal values or religion & 13 & 5.0 & Morals, values, religion, kindness, or broader worldview. \\
    Life experience or upbringing & 9 & 3.4 & Upbringing, lived experience, or breadth of background. \\
    Age or life stage & 9 & 3.4 & Age, generation, grandchildren, or future-oriented life-stage considerations. \\
    Critique of displayed demographics & 7 & 2.7 & Displayed demographic attributes described as irrelevant, overemphasized, or poorly specified. \\
    Diet & 6 & 2.3 & Veganism, vegetarianism, or other dietary practices. \\
    Additional identity characteristics & 5 & 1.9 & Sexuality, neurotype, immigration background, or other identity details. \\
    Institutional experience or conflicts & 5 & 1.9 & Prior assembly or public-affairs experience, willingness to serve, elected office, or vested interests. \\
    Transport behavior & 3 & 1.1 & Driving, flying, or reliance on particular modes of transport. \\
    Legal or safety background & 2 & 0.8 & Criminal history, victimization, or related legal and safety experience. \\
    \bottomrule
  \end{tabular}
\end{table}
\FloatBarrier

\paragraph{Interpretation and limitations.}
The two free-text items are not directly comparable: Survey~1 requested general
feedback, whereas Survey~2 explicitly elicited omitted attributes. Moreover,
the Survey~2 item was answered by less than half of the cleaned sample, and its
counts measure mentions rather than the strength or priority of a preference.
The Survey~1 comments are therefore illustrative, and the Survey~2 themes
should be treated as nominations for future study rather than evidence that
each belongs in a quota schema.

\section{Omitted-Attribute Analysis Details}
\label{sec:appendix-omitted-attributes}

\subsection{Observed Pool and Party--Concern Association}

The omitted-attribute experiment uses one record for each of the 1,151
respondents in Survey~1. The general-purpose and climate-focused rows in that
survey refer to the same people, so we use the general-purpose row to obtain one
copy of each respondent's seven demographic attributes, political alignment,
and self-reported climate concern. Climate concern was not displayed in the
general-purpose conjoint profiles; here it is used only as a panel-composition
outcome or as an explicit quota in the direct-control benchmark. The reference
distribution is therefore the observed Survey~1 pool, not the UK population:
3.74\% not at all concerned, 14.94\% not very concerned, 43.53\% fairly
concerned, and 37.79\% very concerned.

\begin{table}[hbt!]
  \centering
  \small
  \caption{Party-by-climate-concern counts in the cleaned Survey~1 analysis
  pool used for the omitted-attribute experiment ($N=1{,}151$).}
  \label{tab:appendix-party-concern-matrix}
  \begin{tabular}{lrrrr}
    \toprule
    & Not at all & Not very & Fairly & Very \\
    \midrule
    Conservative & 4 & 41 & 84 & 47 \\
    Green & 1 & 1 & 47 & 107 \\
    Labour & 4 & 34 & 158 & 159 \\
    Liberal Democrat & 0 & 10 & 67 & 44 \\
    Other/None & 8 & 30 & 58 & 36 \\
    Plaid Cymru & 0 & 1 & 3 & 6 \\
    Reform UK & 23 & 54 & 69 & 26 \\
    SNP & 3 & 1 & 15 & 10 \\
    \bottomrule
  \end{tabular}
\end{table}

We compute the reported party--concern association from the $8\times4$
contingency table in Table~\ref{tab:appendix-party-concern-matrix}. For a table
with $r$ rows, $c$ columns, Pearson statistic $\chi^2$, and sample size $N$, we
use the conventional uncorrected statistic
\[
 V=\sqrt{\frac{\chi^2}{N\min(r-1,c-1)}}.
\]
Here $\chi^2=234.33$, $N=1{,}151$, $r=8$, and $c=4$, giving
$V=0.261$ ($21$ degrees of freedom, $p<10^{-16}$). This aggregate association
does not imply that balancing party will balance concern in every feasible
panel; the panel experiment tests that implication directly.

\subsection{Primary LEXIMIN Experiment}

We compare four sampling regimes for panels of size 40:
\begin{enumerate}
  \item Uniform sampling draws panels without replacement and imposes no quota
  constraints.
  \item Demographic stratification imposes marginal quotas on gender, age, area
  type, ethnicity, disability, education, and region.
  \item Demographic and political stratification imposes the seven demographic
  quotas together with a marginal quota on political alignment.
  \item Full stratification imposes the seven demographic quotas together with
  marginal quotas on political alignment and climate concern.
\end{enumerate}
For each constrained attribute, its observed pool share is multiplied by the
panel size, and each result is rounded down. Any remaining seats are assigned
to the categories with the largest fractional remainders, ensuring that the
targets sum to the panel size. Lower and upper bounds are one seat below and
above each target, truncated to feasible count limits. For example, the concern
targets in the direct-control regime are 17 fairly concerned, 6 not very
concerned, 2 not at all concerned, and 15 very concerned.

\paragraph{Released implementation.}
The constrained regimes invoke
\texttt{find\_distribution\_leximin} from the released replication source of
\citet{flanigan2021fair}, rather than reimplementing its control flow. The
source was byte-verified against released commit \texttt{6ad2d5fb}; only its optimization-model
interface was mapped from Gurobi to SciPy/HiGHS. We pass the study-specific
quota bounds, the 1,151 people, panel size 40, no household constraints, and no
preselected people. LEXIMIN constructs a distribution over quota-feasible
committees that lexicographically maximizes individual inclusion probabilities,
starting from the least included individual.

For each quota schema, the algorithm returns a representation containing 3,453
generated committees; committees assigned zero probability are retained in the
stored output for auditability. The individual inclusion probabilities are
numerically equal to
$40/1{,}151=0.03475$. We verified that distribution probabilities sum to one,
every positively supported committee satisfies every lower and upper bound,
and the maximum observed quota violation is zero seats.
For reproducibility, we run with Python hash seed zero and sort the
returned committees canonically before drawing 250 panels from each distribution
with seed 20260719. This is
the released LEXIMIN distribution; it should not be interpreted as the uniform
distribution over all quota-feasible panels.

\begin{table}[hbt!]
  \centering
  \small
  \caption{Climate-concern
  total-variation error for 250 size-40 panels drawn from the Survey~1 pool.
  Quota rows use LEXIMIN; percentiles are empirical across the sampled panels.}
  \label{tab:appendix-sampling-primary}
  \begin{tabular}{lrrr}
    \toprule
    Sampling regime & Mean & Median & 5th--95th pct. \\
    \midrule
    Uniform random & .0939 & .0868 & .0310--.1724 \\
    Demographic quotas & .0945
      & .0897 & .0279--.1853 \\
    Demographic + party & .0910
      & .0868 & .0372--.1647 \\
    Demographic + party + concern & .0353
      & .0368 & .0153--.0597 \\
    \bottomrule
  \end{tabular}
\end{table}

Table~\ref{tab:appendix-sampling-primary} reports the primary result.
Adding party to demographic
quotas changes mean concern error by $-0.0035$; the party-augmented mean is
$0.0029$ below uniform sampling despite the observed party--concern association.
Direct concern quotas reduce mean TV by $0.0587$, approximately 62\% relative
to uniform sampling. These comparisons concern the
observed Survey~1 pool and the tested LEXIMIN distributions; they are not a
universal claim about every population or every distribution over feasible
panels.

\subsection{Panel-Size, Tolerance, and Quota Robustness}

We additionally vary panel size $k\in\{20,40,60,80\}$ and quota tolerance
between the exact integer targets described above, $\pm1$ seat, and $\pm2$
seats. These auxiliary runs use random linear objectives to obtain feasible
panels, rather than LEXIMIN, so they test whether the qualitative result depends
on one panel-construction procedure. They do not sample uniformly from all
feasible panels. For each plotted sampling regime, we generate 50
panels per cell under exact, $\pm1$-seat, and $\pm2$-seat quotas. Uniform
benchmarks are computed separately at each panel size.

\begin{figure*}[hbt!]
  \centering
  \includegraphics[width=\textwidth]{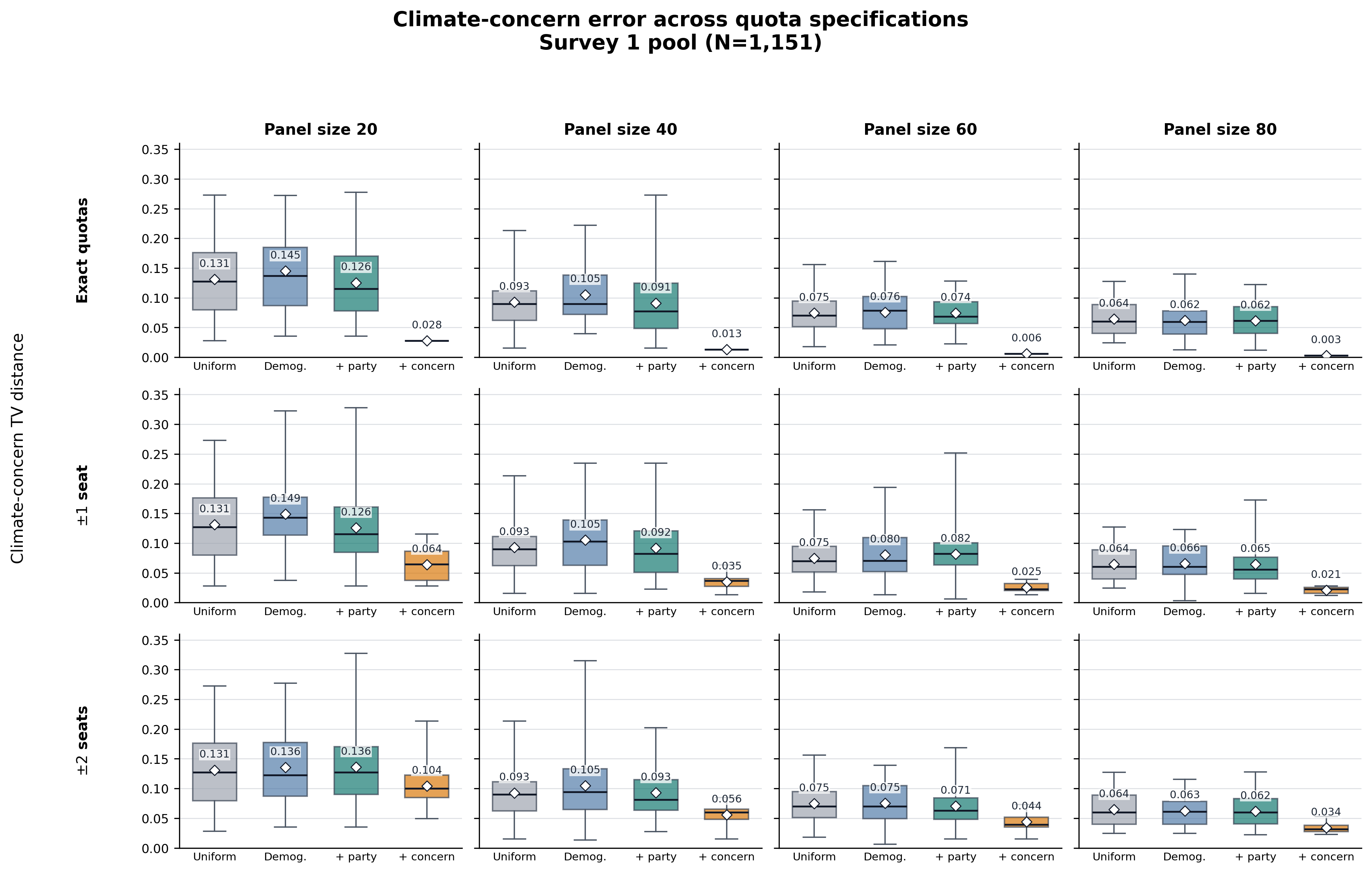}
  \caption{Climate-concern total-variation distributions in the
  Survey~1 robustness experiment. Columns give panel size and rows give exact,
  $\pm1$-seat, and $\pm2$-seat quota tolerances. Within each panel, the four
  boxplots compare uniform sampling, demographic quotas, demographic-plus-party
  quotas, and direct concern quotas across 50 generated panels. Boxes span the
  first and third quartiles, horizontal lines mark medians, whiskers show the
  minimum and maximum, and white diamonds and labels report means.}
  \Description{A three-by-four grid of boxplots comparing four panel
  sampling regimes. Direct concern quotas generally produce lower
  climate-concern total-variation distance than uniform, demographic, or
  demographic-plus-party sampling, particularly with exact or one-seat quota
  tolerances.}
  \label{fig:appendix-sampling-robustness-distributions}
\end{figure*}

Figure~\ref{fig:appendix-sampling-robustness-distributions}
presents box-and-whisker summaries of the generated panels in each robustness
cell. These summaries
describe simulation variation across generated panels, not population
confidence intervals: their broad overlap shows that demographic and
demographic-plus-party panels have error distributions similar to uniform
sampling, whereas directly constrained concern panels are concentrated at
lower error, especially under exact and one-seat-tolerance quotas.
Direct concern constraints yield lower mean concern error than the corresponding
uniform benchmark in every feasible grid cell. Demographic and
demographic-plus-party quotas sometimes improve and sometimes worsen mean TV,
with no consistent advantage across panel sizes and tolerances. The robustness
analysis therefore supports the narrow
interpretation of the primary result: in this pool, the tested demographic and
party quotas do not reliably proxy for an omitted concern margin, while directly
constraining concern controls that margin by construction.

\section{Predictive Models and Results Across Tasks}
\label{sec:metric-learning}
\label{sec:appendix-predictive-comparison}

This appendix gives implementation details for the five predictive models compared in
the main predictive section. It also presents the weighted-$\ell_2$ variant, robustness
checks, and diagnostics that supplement the all-task comparison in
Table~\ref{tab:lp-results}.

\subsection{Common Evaluation Design}

All models use direct pairwise choices and the same 20 reproducible 85/15
participant-level splits. No participant contributes responses to both
partitions of a split. The General Assembly and Climate Assembly
each contain 5,755 complete comparisons from 1,151 participants. The both-cues
Climate Assembly contains 5,760 complete comparisons from 576 participants.
All preprocessing that uses outcomes---including the AMCE ordering and fitted
model coefficients---is recomputed using training respondents only. Exact
prediction ties receive half credit. The dispersion reported below is the
standard deviation across overlapping splits of the same respondent sample,
not a population-level confidence interval.

\subsection{Formal Model Specifications}
\label{sec:appendix-formal-models}

Let $\vec p_i$ denote the respondent profile in comparison $i$, and let
$\vec c_{i1}$ and $\vec c_{i2}$ denote the two candidate profiles. We code
$y_i=1$ when $\vec c_{i1}$ is chosen and $y_i=-1$ when $\vec c_{i2}$ is
chosen. Every model produces an antisymmetric score
$s_i=s(\vec p_i,\vec c_{i1},\vec c_{i2})$. A positive score predicts
$\vec c_{i1}$, a negative score predicts $\vec c_{i2}$, and a zero score is a
symmetric tie worth one-half in the accuracy calculation.

For the three heuristic models, let
\[
  m_{ijh}=\mathbf 1\{p_{ih}=c_{ijh}\}
\]
indicate whether candidate $j\in\{1,2\}$ matches respondent $i$ on attribute
$h$. Each heuristic produces a raw score $r_i$, described below.

\paragraph{Model 1: attribute match count.}
Define
\[
  M(\vec p_i,\vec c_{ij})
  =
  \sum_{h\in\mathcal A}m_{ijh},
  \qquad
  r_i^{\mathrm{count}}
  =
  M(\vec p_i,\vec c_{i1})-M(\vec p_i,\vec c_{i2}).
\]
The candidate with more exact matches is selected; equal match counts yield
$r_i^{\mathrm{count}}=0$. Every displayed attribute has equal weight,
ordinal gap size is ignored, and no numerical choice-rule coefficient is
fitted.

\paragraph{Training-only mismatch AMCEs.}
The next two models use an ordering learned separately within each training
split. We expand every comparison into two candidate rows and let
$A_{ij}$ indicate whether candidate $j$ was selected. For respondent $u$,
attribute $h$, and mismatch state $m\in\{0,1\}$, let
$\overline A_{uhm}$ be that respondent's mean candidate-selection indicator
among candidate rows with
$\mathbf 1\{p_{uh}\ne c_{ijh}\}=m$. If $\mathcal U_h$ contains the training
respondents observed in both mismatch states, the training AMCE is
\[
  \widehat\tau_h
  =
  \frac{1}{|\mathcal U_h|}
  \sum_{u\in\mathcal U_h}
  \left(\overline A_{uh1}-\overline A_{uh0}\right).
\]
Attributes are ranked by decreasing $|\widehat\tau_h|$, with ties resolved by
their original display order. If no training respondent appears in both
mismatch states, we set $\widehat\tau_h=0$. No test outcomes enter either the
estimates or the ordering.

\paragraph{Model 2: most-important cue.}
Let
\[
  h^\star=\arg\max_{h\in\mathcal A}|\widehat\tau_h|,
  \qquad
  r_i^{\mathrm{cue}}=m_{i1h^\star}-m_{i2h^\star}.
\]
The model selects a candidate only when exactly that candidate matches the
respondent on the focal attribute; if both or neither candidate matches, its
raw score is zero. The selected cue is party in every General Assembly split,
climate concern in every Climate Assembly split, and party in
every both-cues split. As a secondary diagnostic, using climate concern instead
of party in the both-cues task yields train and test accuracies of .575 and
.572.

\paragraph{Model 3: AMCE-ordered match tree.}
Let $\pi=(\pi_1,\ldots,\pi_H)$ be the training-derived attribute order and
define
\[
  k_i
  =
  \min\{k:m_{i1\pi_k}\ne m_{i2\pi_k}\}.
\]
If this set is nonempty, the raw score is
\[
  r_i^{\mathrm{tree}}
  =
  m_{i1\pi_{k_i}}-m_{i2\pi_{k_i}};
\]
otherwise it is zero. Thus, the tree stops at the first ordered attribute on
which exactly one candidate matches the respondent and never consults a
lower-ranked attribute after that point. The average Climate
Assembly order is climate concern, region, age, education, ethnicity, gender,
area type, and disability status.

\paragraph{Model 4: weighted $\ell_1$ metric.}
Let $\mathcal O$ contain the ordinal attributes---age group, education level,
and climate concern---and let $o_h(v)\in\{1,\ldots,K_h\}$ be the ordered index
of value $v$ for $h\in\mathcal O$. For respondent value $p_h$ and candidate
value $c_h$, define the attribute distance 
\[
  \delta_h(p_h,c_h)
  =
  \begin{cases}
    \dfrac{|o_h(p_h)-o_h(c_h)|}{K_h-1},
      & h\in\mathcal O,\\[2mm]
    \mathbf 1\{p_h\ne c_h\},
      & h\notin\mathcal O.
  \end{cases}
\]
Thus, $\delta_h\in[0,1]$: a match has distance zero, an ordinal mismatch grows
with the normalized category gap, and any mismatch on an unordered attribute
has distance one. For example, the four age categories are ordered as
16--29, 30--44, 45--60, and 60+; for a 16--29 respondent, a 30--44 candidate
has age distance $1/3$, whereas a 60+ candidate has age distance $3/3=1$. Let $\vec\Delta(\vec p,\vec c)$ collect these distances, so
$d_{\vec w}(\vec p,\vec c)=\vec w^\top\vec\Delta(\vec p,\vec c)$. For
comparison $i$, define
\[
  \vec\phi_i
  =
  \vec\Delta(\vec p_i,\vec c_{i2})
  -
  \vec\Delta(\vec p_i,\vec c_{i1}),
  \qquad
  s_i^{\ell_1}=\vec\phi_i^\top\vec w.
\]
We estimate one nonnegative weight per displayed attribute:
\[
 \widehat{\vec w}
 =
 \arg\min_{\vec w\geq 0}
 \left\{
 \frac{1}{N_{\mathrm{tr}}}\sum_{i\in\mathrm{tr}}
 \max\!\left(0,1-y_i\vec\phi_i^\top\vec w\right)
 +
 \frac{0.01}{2}\|\vec w\|_2^2
 \right\}.
\]
This is the reduced hinge-loss form of the margin-one quadratic program with
one nonnegative slack variable per training comparison. We solve the
nonnegative constrained problem from both all-zero and all-one initial weights
and retain the lower converged objective; all 60 fitted splits converge.
Prediction compares the two distances directly:
$\vec c_{i1}$ is selected when $\vec\phi_i^\top\widehat{\vec w}>0$,
$\vec c_{i2}$ when it is negative, and zero is a tie. The model has $H$
fitted choice-rule parameters: eight in the General and Climate
Assemblies and nine in the both-cues task.

\paragraph{Model 5: Weighted $\ell_2$ metric.}
We also fit a Euclidean variant that squares each attribute distance before
aggregating, defining
\[
\phi_{ih}^{\ell_2} = \delta_h(p_{ih}, c_{i2h})^2 - \delta_h(p_{ih}, c_{i1h})^2,
\qquad
s_i^{\ell_2} = {\phi_i^{\ell_2}}^\top\mathbf{w}.
\]
The score is again linear in $\mathbf{w}$, so the estimator is the same
nonnegative hinge-loss program with the same penalty, initializations, and
respondent-level splits as weighted $\ell_1$.
\paragraph{Model 6: conditional Bradley--Terry.}
This model belongs to the class of covariate-extended Bradley--Terry models,
which incorporate subject attributes, candidate attributes, and interactions
between them
\citep{dittrich1998subject}. Equivalently, it is a binary conditional-logit
utility model with difference-coded candidate attributes and
respondent--candidate interactions. We use \emph{conditional BT} as shorthand
for the block-restricted specification below.

All categorical values are reference-category one-hot encoded using the same
outcome-free category vocabulary for respondents and candidates. Write the
encoding in attribute blocks as
$\vec x=(\vec x^{(1)},\ldots,\vec x^{(H)})$, let $\vec z_i$ be the encoding of
respondent $\vec p_i$, and define
\[
  \vec d_i=\vec x_{c_{i1}}-\vec x_{c_{i2}}.
\]
For an attribute with $K_h$ categories, its block has
$q_h=K_h-1$ coordinates. Our block-conditional specification is
\[
  s_i^{\mathrm{BT}}
  =
  \vec\beta^\top\vec d_i
  +
  \sum_{h=1}^{H}
  {\vec z_i^{(h)}}^\top W_h\vec d_i^{(h)}.
\]
Equivalently, the logistic-regression design row is
\[
  \left[
    \vec d_i;\,
    \vec z_i^{(1)}\otimes\vec d_i^{(1)};\,
    \ldots;\,
    \vec z_i^{(H)}\otimes\vec d_i^{(H)}
  \right].
\]
The first term estimates average candidate-value utilities. Each
$q_h\times q_h$ matrix $W_h$ then allows those utilities to vary with the
respondent's value of the same attribute. Placing the $W_h$ blocks on the
diagonal of $W$ retains every same-attribute interaction but sets all
cross-attribute interactions---such as respondent age by candidate region or
respondent gender by candidate education---to zero. This restriction preserves
value-specific within-attribute heterogeneity without fitting the full outer
product of every respondent and candidate indicator.

The number of fitted coefficients is
\[
  Q+\sum_{h=1}^{H}q_h^2,
  \qquad
  Q=\sum_{h=1}^{H}q_h.
\]
In the  Climate Assembly, the blocks for gender, age group, area
type, ethnicity, disability status, education, region, and climate concern
have dimensions
\[
  (q_h)_{h=1}^{8}=(2,3,2,4,2,5,11,3).
\]
Consequently, $Q=32$ candidate main-effect coefficients and
$\sum_hq_h^2=192$ interaction coefficients give 224 coefficients in total.
The corresponding totals are 268 for the General Assembly and 280 for the
both-cues task.

Writing $\vec\theta$ for the concatenated coefficients, we minimize an
L2-penalized logistic objective of the form
\[
  \sum_{i\in\mathrm{tr}}
  \log\!\left(1+\exp[-y_i s_i^{\mathrm{BT}}(\vec\theta)]\right)
  +
  \frac{\lambda_{\mathrm{BT}}}{2}\|\vec\theta\|_2^2.
\]
We use the scikit-learn inverse-regularization setting $C=1$, L-BFGS,
tolerance $10^{-8}$, and at most 5,000 iterations. The model has no intercept,
so exchanging $\vec c_{i1}$ and $\vec c_{i2}$ negates $\vec d_i$ and the score,
ensuring
\[
  \Pr(\vec c_{i1}\succ\vec c_{i2}\mid\vec p_i)
  =
  1-\Pr(\vec c_{i2}\succ\vec c_{i1}\mid\vec p_i).
\]
It also contains no participant identifiers: a test respondent enters only
through the observed one-hot profile $\vec z_i$.

\subsection{Results Across All Task Environments}

\begin{table}[!ht]
  \centering
  \small
  \setlength{\tabcolsep}{5pt}
  \caption{Weighted-$\ell_1$ and weighted-$\ell_2$ predictive
  accuracy across task environments (mean $\pm$ s.d.\ over 20 shared
  participant-level splits). The main text reports the complete five-model
  comparison.}
  \label{tab:appendix-predictive-all}
  \begin{tabular}{lrrrr}
    \toprule
    & \multicolumn{2}{c}{Weighted $\ell_1$}
    & \multicolumn{2}{c}{Weighted $\ell_2$} \\
    \cmidrule(lr){2-3}\cmidrule(lr){4-5}
    Task & Train acc. & Test acc. & Train acc. & Test acc. \\
    \midrule
    General-purpose
      & $.732\pm.003$ & $.725\pm.015$
      & $.729\pm.003$ & $.724\pm.016$ \\
    Climate, concern only
      & $.760\pm.003$ & $.758\pm.018$
      & $.740\pm.003$ & $.736\pm.018$ \\
    Climate, both cues
      & $.692\pm.003$ & $.687\pm.015$
      & $.687\pm.003$ & $.682\pm.014$ \\
    \bottomrule
  \end{tabular}
\end{table}

\begin{figure}[!ht]
  \centering
  \includegraphics[width=\textwidth]
    {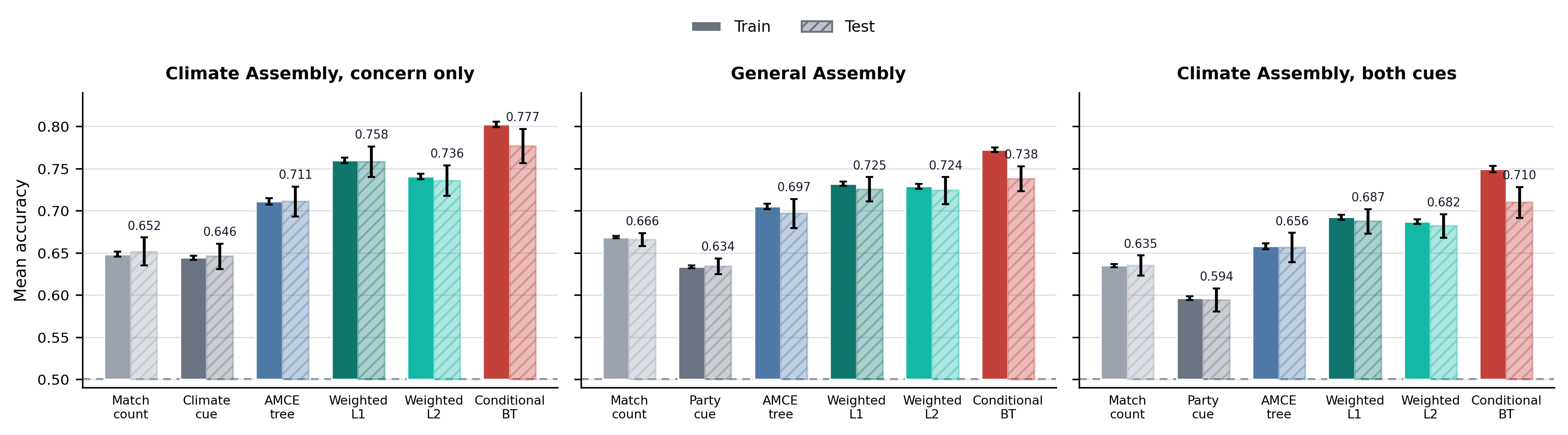}
  \caption{Mean training and held-out accuracy across the three task
  environments. Solid bars show training accuracy and hatched bars show test
  accuracy; numbers above the hatched bars report mean test accuracy to three
  decimal places. Error
  bars give one standard deviation over the 20 shared participant-level
  splits. Dashed horizontal lines mark chance accuracy.}
  \Description{Three grouped bar charts compare training and test accuracy for
  attribute match count, the focal-cue heuristic, the AMCE-ordered tree,
  weighted L1, and conditional Bradley--Terry. The train and test bars are
  nearly equal for weighted L1. Conditional Bradley--Terry has the highest
  test accuracy but a visibly larger train--test gap in every task. Numerical
  labels above the hatched bars give the corresponding mean test accuracies.}
  \label{fig:appendix-predictive-train-test}
\end{figure}

Table~\ref{tab:lp-results} in the main text reports the
five-model comparison across all tasks. Figure~\ref{fig:appendix-predictive-train-test}
visualizes the same ordering and the corresponding train--test gaps.
Weighted $\ell_2$ does not improve on weighted $\ell_1$ in any of the three
tasks. Mean held-out accuracy is lower by 0.1 percentage points in the General
Assembly, 2.2 points in the Climate Assembly, and 0.5 points in
the both-cues task. The split-level standard
deviations are descriptive and, because the splits overlap, do not provide an
inferential test of the difference between model families. We retain the
$\ell_1$ aggregation because its mean accuracy is no lower in any task and its
additive distance is easier to interpret, not because these results establish
that $\ell_1$ is generally superior.

The paired bars also reveal systematically different train--test gaps.
Conditional BT's mean training accuracy exceeds its test accuracy by 3.4
points in the General Assembly, 2.6 points in the Climate
Assembly, and 4.0 points in the both-cues task. The corresponding gaps for
weighted $\ell_1$ are only 0.7, 0.2, and 0.5 points. Given the much larger number of conditional-BT
coefficients, its larger train--test gap is consistent with possible
overfitting, although the gap alone does not identify its source. Conditional
BT nevertheless retains the highest mean held-out accuracy in every task; the
generalization concern therefore qualifies, but does not reverse, the observed
test-set ordering.

\subsection{Complementary Predictive Information from Party and
Climate Concern}
\label{sec:appendix-cue-complementarity}

The main model comparison holds the information supplied to each model
fixed. Survey~2 serves a different purpose: because political affiliation and
climate concern appear together, it allows us to ask whether they provide
overlapping or complementary predictive information. We fit weighted
$\ell_1$ and conditional BT to four deliberately chosen attribute sets: the
seven demographic and geographic attributes alone; those attributes plus
party; those attributes plus concern; and all nine attributes. For every
specification, respondent and candidate profiles are restricted to the same
attribute set. The models use the same 5,760 choices from 576 respondents, the
same 20 participant-level splits, and the same estimation settings as in the
main comparison. Only training respondents enter the fitted coefficients;
weighted $\ell_1$'s auxiliary probability scale is also fitted on the training
data.

The Survey~2 design provides independent variation for this
comparison. Although sampled party and concern values retain their smoothed
empirical association, the candidate's party-match and concern-match
indicators are drawn independently, each with probability one-half.
Consequently, the four match--mismatch combinations are balanced by design,
allowing each cue to vary while the other is held fixed. Weighted $\ell_1$
uses 7, 8, 8, and 9 fitted weights across the four information sets. The
corresponding conditional-BT specifications use 212, 268, 224, and 280
coefficients because party has more categorical levels than concern.

\begin{table}[!ht]
  \centering
  \small
  \setlength{\tabcolsep}{3.5pt}
  \caption{Predictive performance by attribute set in the both-cues Climate
  Assembly (mean $\pm$ s.d.\ over 20 shared participant-level splits). All
  specifications are evaluated on identical test observations.}
  \label{tab:appendix-both-cues-information}
  \begin{tabular}{llrrrr}
    \toprule
    Model & Attribute set & \# Parameters & Train acc. & Test acc. & Test log loss \\
    \midrule
    Weighted $\ell_1$
      & Demographics & 7 & $.604\pm.003$ & $.602\pm.014$ & $.666\pm.008$ \\
      & Demographics + party & 8 & $.658\pm.003$ & $.655\pm.013$ & $.629\pm.012$ \\
      & Demographics + concern & 8 & $.653\pm.004$ & $.646\pm.019$ & $.629\pm.015$ \\
      & Demographics + both & 9 & $.692\pm.003$ & $\mathbf{.687\pm.015}$ & $\mathbf{.589\pm.016}$ \\
    \midrule
    Conditional BT
      & Demographics & 212 & $.638\pm.004$ & $.596\pm.017$ & $.676\pm.010$ \\
      & Demographics + party & 268 & $.711\pm.003$ & $.670\pm.015$ & $.611\pm.018$ \\
      & Demographics + concern & 224 & $.697\pm.003$ & $.658\pm.016$ & $.622\pm.015$ \\
      & Demographics + both & 280 & $.749\pm.004$ & $\mathbf{.710\pm.018}$ & $\mathbf{.561\pm.022}$ \\
    \bottomrule
  \end{tabular}
\end{table}

\begin{figure}[!ht]
  \centering
  \includegraphics[width=\textwidth]
    {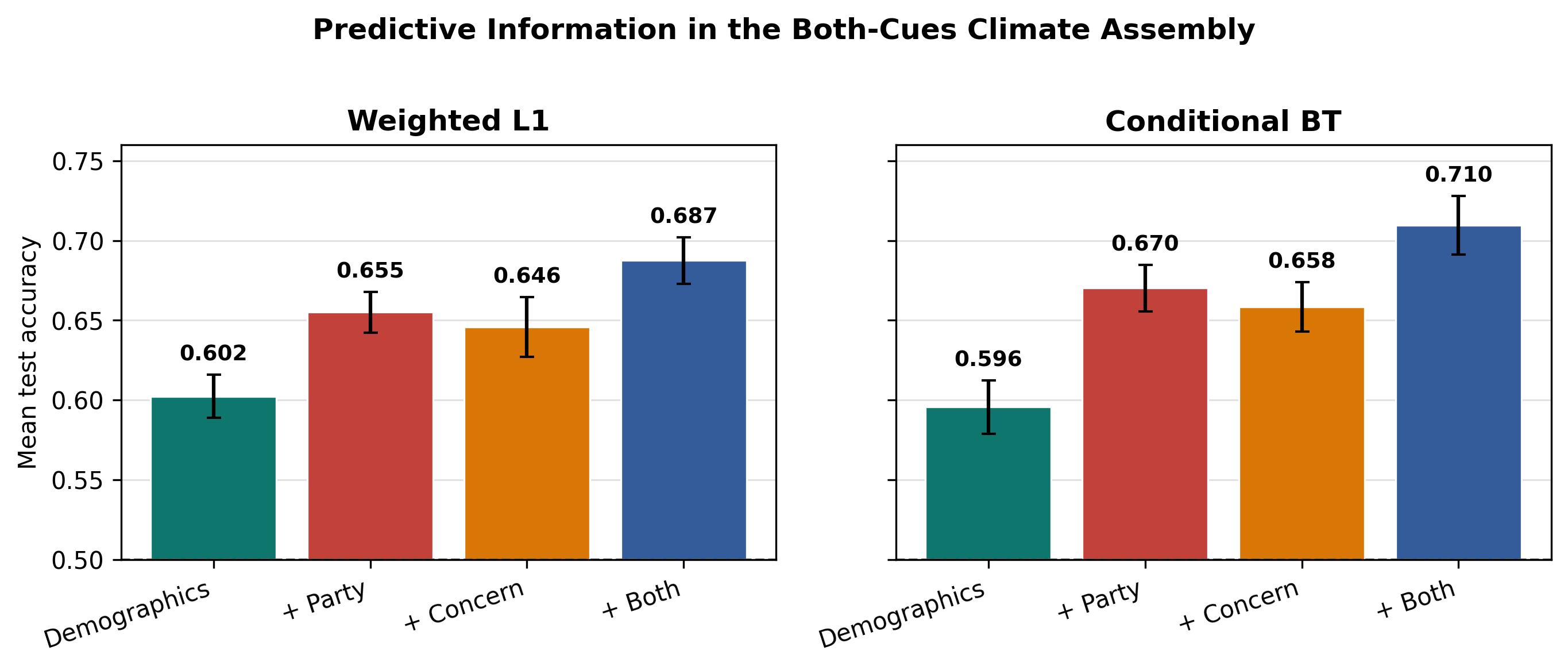}
  \caption{Held-out accuracy by attribute set in the both-cues Climate
  Assembly. Bars report means over 20 shared participant-level splits, error
  bars show one standard deviation, and numerical labels report the means.
  Every model is evaluated on the same test observations in a given split.}
  \Description{Two grouped bar charts show held-out accuracy for weighted L1
  and conditional Bradley--Terry using demographics alone, demographics plus
  party, demographics plus climate concern, and demographics plus both.
  Accuracy is highest when both party and concern are included for each model.}
  \label{fig:appendix-both-cues-complementarity}
\end{figure}

The descriptive ordering is unambiguous
(Table~\ref{tab:appendix-both-cues-information} and
Figure~\ref{fig:appendix-both-cues-complementarity}). Either cue improves on
demographics alone, and the specification containing both performs best.
For weighted $\ell_1$, test accuracy increases from .655 with party and .646
with concern to .687 with both. Conditional BT rises from .670 and .658,
respectively, to .710. Held-out log loss follows the same ordering, so the
result is not an artifact of thresholding probabilities at one-half.

We assess each cue's incremental contribution by pairing predictions
on the same test observations. For each respondent, we first average the
repeated held-out differences across every split in which that respondent
appears, then form 20,000 percentile-bootstrap samples at the respondent
level. Of the 576 respondents, 554 appear in a test set at least once across
the 20 splits. Table~\ref{tab:appendix-both-cues-contrasts} reports the
resulting paired summaries.

\begin{table}[!ht]
  \centering
  \small
  \setlength{\tabcolsep}{5pt}
  \caption{Incremental predictive contribution of each cue. Accuracy gains
  are percentage points; positive log-loss reductions indicate improvement.
  Brackets give 95\% paired participant-bootstrap intervals.}
  \label{tab:appendix-both-cues-contrasts}
  \begin{tabular}{llrr}
    \toprule
    Model & Added information & Accuracy gain & Log-loss reduction \\
    \midrule
    Weighted $\ell_1$
      & Concern, given party & $+3.2\ [2.1,4.3]$ & $.040\ [.032,.049]$ \\
      & Party, given concern & $+4.0\ [2.8,5.2]$ & $.042\ [.032,.051]$ \\
    Conditional BT
      & Concern, given party & $+4.0\ [2.9,5.1]$ & $.050\ [.039,.062]$ \\
      & Party, given concern & $+5.2\ [4.1,6.3]$ & $.063\ [.052,.075]$ \\
    \bottomrule
  \end{tabular}
\end{table}

\FloatBarrier

All four accuracy intervals and all four log-loss intervals exclude
zero. Thus, concern contributes predictive information after party is
included, and party contributes after concern is included. The same pattern
under both model families indicates that it is not peculiar to one functional
form. These intervals summarize paired predictive variation in this
experimental sample rather than population-level causal uncertainty. The
result has a correspondingly focused interpretation: within this experiment,
neither political affiliation nor climate concern subsumes the predictive
information supplied by the other.

\subsection{Participant-Level Predictive Consistency}

Aggregate accuracy pools choices across respondents and can therefore conceal
whether errors are broadly distributed or concentrated among particular
respondents. To examine this, we count the number of each test respondent's
choices that a model predicts correctly. Let $s_{vim}^{(r)}$ be model $m$'s
score favoring $\vec c_1$ over $\vec c_2$ in choice $i$ of test respondent $v$
under split $r$, and let $y_{vi}=1$ when $\vec c_1$ was chosen. We define
strict participant-level consistency as
\[
  K_{vm}^{(r)}
  =
  \sum_{i=1}^{n_v}
  \mathbf 1\!\left\{
    (2y_{vi}-1)s_{vim}^{(r)}>0
  \right\}.
\]
For weighted $\ell_1$, the score is
$d_{\vec w}(\vec p,\vec c_2)-d_{\vec w}(\vec p,\vec c_1)$; for conditional BT,
it is the fitted utility difference. Thus, $K_{vm}^{(r)}$ is defined
identically for both models. An exact score tie is not strictly consistent and
contributes zero. Such ties constitute at most 0.23\% of test comparisons in
any model--task combination, so this convention has negligible effect on the
distributions. We compute the distribution of $K_{vm}^{(r)}$ within each split
and average the respondent shares over the 20 shared splits.

\begin{figure}[!ht]
  \centering
  \includegraphics[width=\textwidth]
    {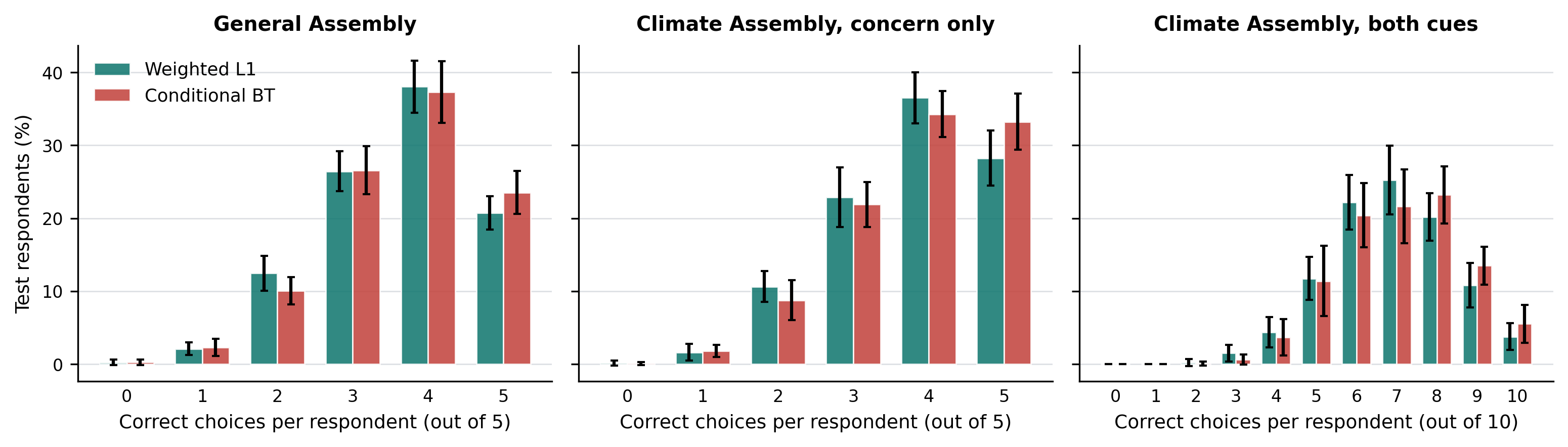}
  \caption{Participant-level predictive consistency for weighted $\ell_1$ and
  conditional BT. The horizontal axis counts choices for which the model
  strictly favors the respondent's observed choice. Bars show the mean
  percentage of test respondents at each count over the 20 shared
  participant-level splits; error bars show one standard deviation. General
  and Climate respondents make five choices each, while
  both-cues Climate respondents make ten. Exact score ties count as
  inconsistent.}
  \Description{Three grouped bar charts show the distribution of correctly
  predicted choices per test respondent for weighted L1 and conditional
  Bradley--Terry in the General Assembly,  Climate Assembly, and
  both-cues Climate Assembly. The two distributions overlap substantially,
  with conditional Bradley--Terry placing more mass modestly on higher correct
  counts.}
  \label{fig:appendix-participant-consistency}
\end{figure}

Conditional BT shifts some mass toward higher consistency in every task, but
the two distributions overlap substantially. In the Climate
Assembly, weighted $\ell_1$ predicts at least four of five choices for 64.7\%
of test respondents and all five for 28.2\%; the corresponding conditional-BT
shares are 67.5\% and 33.2\%. In the General Assembly, the shares with at least
four correct are 58.8\% and 60.8\%, respectively. In the both-cues task,
conditional BT's advantage is more visible at the upper end: it predicts at
least eight of ten choices for 42.2\% of test respondents, compared with 34.8\%
for weighted $\ell_1$. Perfect prediction remains uncommon there (5.5\% versus
3.8\%). The participant-level results therefore reinforce the aggregate
comparison: conditional BT improves prediction, but it does not produce a
qualitatively different distribution of model fit across respondents.

\subsection{Learned Weighted-$\ell_1$ Profiles}

Table~\ref{tab:appendix-l1-weights} reports each attribute's
mean share of learned weighted-$\ell_1$ importance together with the
corresponding normalized absolute-AMCE profile. For presentation only, the
learned weights are normalized to sum to one within each fitted split before
averaging. The AMCE profile divides each attribute's absolute primary
AMCE by the sum of the absolute AMCEs within that task. Political affiliation has the largest
mean weight in the General Assembly, climate concern has the largest weight in
the  Climate Assembly, and concern and party receive the two
largest weights when both are shown. The remaining weight is distributed
across the demographic attributes rather than collapsing onto the dominant
cue.

\begin{table}[!ht]
  \centering
  \scriptsize
  \setlength{\tabcolsep}{3pt}
  \caption{Mean
  normalized weighted-$\ell_1$ coefficients across the 20 training splits and
  normalized absolute primary AMCEs. Dashes indicate attributes not displayed
  in that task.}
  \label{tab:appendix-l1-weights}
  \resizebox{\textwidth}{!}{%
  \begin{tabular}{lrrrrrrrrr}
    \toprule
    Task and quantity & Gender & Age & Area & Ethnicity & Disability & Education
      & Region & Party & Concern \\
    \midrule
    General, learned $\ell_1$
      & .071 & .221 & .041 & .072 & .027 & .163 & .125 & .278 & -- \\
    General, normalized $|\mathrm{AMCE}|$
      & {.092} & {.141} & {.047} & {.081}
      & {.023} & {.111} & {.141} & {.363} & -- \\
    \addlinespace[2pt]
    Climate, learned $\ell_1$
      & .054 & .144 & .043 & .059 & .015 & .121 & .104 & -- & .460 \\
    Climate, normalized $|\mathrm{AMCE}|$
      & {.078} & {.130} & {.065} & {.085}
      & {.022} & {.090} & {.150} & -- & {.380} \\
    \addlinespace[2pt]
    Climate both-cues, learned $\ell_1$
      & .027 & .142 & .011 & .051 & .021 & .134 & .082 & .191 & .340 \\
    Climate both-cues, normalized $|\mathrm{AMCE}|$
      & {.046} & {.111} & {.020} & {.076}
      & {.026} & {.106} & {.118} & {.278} & {.218} \\
    \bottomrule
  \end{tabular}
  }
\end{table}
\FloatBarrier

The two profiles agree on the dominant cue in the General Assembly (party)
and the Climate Assembly (climate concern). In the both-cues task, however,
normalized absolute AMCE ranks party above concern (.278 versus .218),
whereas the metric model reverses them (.191 versus .340); the metric model
also assigns relatively more importance to the ordinal age and education
attributes than the AMCE profile does.

We believe this reversal reflects how the attributes are encoded rather
than a genuine disagreement about which matters more. Party is coded as
flat binary distance (mismatch $=1$, match $=0$) in both models, so AMCE
and the weighted-$\ell_1$ weight describe the same quantity for party ---
the average cost of any mismatch. Climate concern, age, and education, by
contrast, are ordinal in weighted-$\ell_1$ (distance scaled by gap size),
so their fitted weights are per-unit-distance slopes rather than average
mismatch costs, while AMCE still averages ``any mismatch vs.\ match''
uniformly across near and far mismatches. Since
Figure~\ref{fig:app-heatmap-climate-concern}
shows
that concern's penalty is concentrated almost entirely in the far tail ---
adjacent-step mismatches are barely penalized --- AMCE is diluted by the
many weak near-miss draws and ranks concern below party, whereas
weighted-$\ell_1$'s slope must be large to reproduce that steep tail and so
ends up looking larger than party's flat weight, even though the two
quantities are in different units and are not directly comparable.

To test this explanation directly, we refit weighted-$\ell_1$ with climate
concern, age, and education each recoded as binary match/mismatch
variables, matching party's encoding. Under this specification the
reversal disappears for all three attributes: political alignment again
receives the larger weight relative to concern, and age and education fall
back in line with their AMCE-based ranking. This supports encoding
asymmetry, rather than a substantive disagreement between the two
analyses, as the source of the reversal.

\end{document}